\documentclass[a4,10pt,english,twocolumn,aps,prl,superscriptaddress]{revtex4-2}

\usepackage{mathtools}
\usepackage{graphicx} 
\usepackage{xcolor}
\usepackage{adjustbox}
\usepackage{subcaption}
\usepackage[colorlinks = true,
            linkcolor = blue,
            urlcolor  = blue,
            citecolor = blue,
            anchorcolor = blue]{hyperref}
\usepackage[normalem]{ulem}
\usepackage{physics}
\usepackage{appendix}
\usepackage{footnote}
\usepackage{orcidlink}

\newcommand{\pcsadd}{Center for Theoretical Physics of Complex Systems, Institute for Basic Science(IBS), Daejeon 34126, Republic of Korea}
\newcommand{\ustadd}{Basic Science Program, Korea University of Science and Technology (UST), Daejeon 34113, Republic of Korea}

\newcommand{\sect}[1]{\emph{#1}---}

\newcommand{\mop}{\mathcal{O}}
\newcommand{\omo}{\overline{\mop}}
\newcommand{\mot}{\mop(t)}

\newcommand{\alexei}[1]{\textcolor{brown}{#1}}
\newcommand{\bb}[1]{\textcolor{black}{#1}}

\newcommand{\updated}[1]{\textcolor{black}{#1}}

\begin{document}

\title{
Thermalization slowing down of weakly nonintegrable quantum spin dynamics
}

\author{Budhaditya Bhattacharjee\,\orcidlink{0000-0003-1982-1346}}
    \affiliation{\pcsadd}
    \email{budhadityab@ibs.re.kr}

\author{Alexei Andreanov\,\orcidlink{0000-0002-3033-0452}}
    \affiliation{\pcsadd}
    \affiliation{\ustadd}
    \email{aalexei@ibs.re.kr}

\author{Sergej Flach\,\orcidlink{0000-0003-1710-3746}}
    \affiliation{\pcsadd}
    \affiliation{\ustadd}
    \email{sflach@ibs.re.kr}

\date{\today}

\begin{abstract}
    We study thermalization slowing down of a quantum many-body spin system upon approach to two distinct integrability limits.
    Motivated by previous studies of classical systems, we identify two thermalization time scales:
    one quantum Lyapunov time scale is extracted by quantifying operator growth in time on an appropriately defined basis,
    while another ergodization time scale is related to the statistics of fluctuations of the time-evolved operator around its mean value based on the eigenstate thermalization hypothesis.
    Using a paradigmatic Quantum Ising chain we find that both timescales diverge upon approach to integrability.
    We investigate the relative strength of the divergence in the two limits and find that despite significant qualitative differences in the mechanism of integrability breaking, the timescales diverge in a similar fashion.
    This allows us to establish a universality of integrability breaking in quantum spin dynamics.
\end{abstract}

\maketitle

\sect{Introduction}
The study of integrable and chaotic dynamics in quantum systems is an area of active investigation with the goal of explaining the emergence of statistical mechanics in interacting quantum systems, among others. 
Multiple observables have been identified and studied as probes of integrable/chaotic dynamics.
One class of such probes is based on spectral properties of the system~\cite{bohigas1984characterization,berry1977level, zakrewski2023quantum,haake2006quantum,brezin1997spectral,garcia2016spectral2,garcia2017spectral,khramtsov2021spectral} rooted in the Bohigas-Giannoni-Schmit (BGS) and Berry-Tabor conjectures~\cite{bohigas1984characterization,berry1977level} and the expected random matrix-like behavior of quantum systems~\cite{forrester2010log, mehta2004random}.
Another class of observables is based on the operator growth or state evolution under integrable/chaotic Hamiltonians.
Observables such as Out-of-time-ordered-correlation functions (OTOCs) ~\cite{stanford2015many,caputa2016out,roberts2016chaos,shen2016out,cotler2017chaos, cotler2017out,hashimoto2017out,fan2017out,sun2018out,tsuji2018out,deMelloKoch2019spectral,akutagawa2020out,hu2020out}, circuit complexity ~\cite{haferkamp2021linear,magan2018black,chapman2021quantum,rocajerat2023circuit}, operator size~\cite{zhang2022dynamical,agarwal2020emergent,roberts2018operator,zhou2019operator} and Krylov complexity~\cite{parker2019A, barbon2019on,rabinovici2020operator,rabinovici2022krylov,rabinovici2022krylov2, bhattacharjee2022krylov,bhattacharjee2022probing,bhattacharjee2023lanczos} fall into this category. 

Thermalization is a closely related phenomenon to the study of chaos.
It describes late-time physics at equilibrium and leads to the emergence of statistical mechanics. Thermalization is a universal property of non-integrable systems, found in both classical and quantum dynamics.
Specifically, we are interested in the nature of thermalization near integrability, where it is expected to slow down.
This has been explored extensively in classical systems.
One of the key features studied in this respect are the relevant timescales: Lyapunov time, ergodization time, etc. These timescales are obtained by computing different observables and studying their divergence upon approach to integrability~\cite{malishava2022thermalization,malishava2022lyapunov,lando2023thermalization,zhang2024thermalization,danieli2017intermittent,danieli2019dynamical,mithun2018weakly,mithun2019dynamical,bel2005weak,rigol2009breakdown}.

In quantum mechanical systems, the Eigenstate Thermalisation Hypothesis (ETH)~\cite{srednicki1994chaos,deutsch1991quantum,dalessio2015from,magan2016random,deutsch2018eigenstate,brenes2020eigenstate,jafferis2023matrix} is often used to characterize thermalization.
Within the purview of ETH, there are ergodization time-scales (e.g. the Thouless time etc.)~\cite{dalessio2015from,Wang2021eigenstate,dymarsky2022bound,lezama2021equilibration,schiulaz2015dynamics} which have been explored in various systems.
These time scales differ between chaotic and integrable systems, and their exact nature is studied extensively.
Another active direction of investigation involves the notion of Lyapunov-like time scales for quantum systems.
Operator growth serves as a potential path (via OTOCs) to define an appropriate spectrum~\cite{rozenbaum2016lyapunov,hallam2018the,gharibyan2018quantum}.
Similarly, the spectral function is also used~\cite{chan2021spectral}.
In quantum systems without well-defined classical limits (for example, spin-\(1/2\) chains) the quantum Lyapunov spectra behave quite differently from classical spectra and may suffer from definition ambiguities~\cite{gharibyan2018quantum,hallam2018the}.
There are better-defined notions of the \textit{maximum Lyapunov exponent}, which is typically extracted from the growth exponent of an appropriately defined observable.
These include the exponent of the OTOC~\cite{maldacena2015A} and Krylov complexity~\cite{parker2019A,bhattacharjee2022krylov}. 

In this manuscript, we extend the concept of time scales, originally developed for classical networks, to quantum mechanical systems near integrability.
We introduce the notion of quantum networks near integrability and characterize them by studying the dynamics of conserved quantities of the limiting integrable Hamiltonian.
We employ the operator growth approach (using Krylov complexity) to define an appropriate notion of Lyapunov time.
Operator growth is captured through the lens of Krylov complexity, which describes the evolution of an operator on a minimal basis.
We then use ETH principles to extract another time scale, which we coin the ergodization timescale (in analogy to classical systems).
The system that we study is a prototypical 1D quantum Ising spin-\(1/2\) chain.
\updated{Near the integrable limits, the two time scales are compared.
Their behavior is used to identify universal features of integrability breaking, by considering qualitatively different mechanisms of integrability breaking, which we call short and long-range networks by analogy with the classical case~\cite{malishava2022thermalization,malishava2022lyapunov,lando2023thermalization,zhang2024thermalization,danieli2017intermittent,danieli2019dynamical,mithun2018weakly,mithun2019dynamical}.}

\sect{The model}
The prototypical spin system we employ in order to characterize long and short-range networks is the Quantum Ising chain (QIC)~\cite{pfeuty1970the,sachdev2007quantum} given by the following Hamiltonian
\begin{align}
    \label{eq:TFIM}
    H = -\sum_{i = 1}^{N}\left(J \sigma^{z}_i\sigma^{z}_{i+1} + g \sigma^{z}_{i} + h \sigma^{x}_{i}\right)
\end{align}
where \(J, g\) and \(h\) are real numbers describing nearest-neighbour interaction, longitudinal and transverse magnetic field respectively.
The \(\sigma_i^{x, z}\) are Pauli matrices, describing spin-\(1/2\) algebra.
The system is in general non-integrable and has been extensively studied through various probes of quantum chaos~\cite{banuls2011strong,roberts2014localized,arul2005multipartite,lin2018otoc,craps2020lyapunov,nivedita2020spectral}.
In the limits \(g \rightarrow 0\) (Transverse Field Ising Model (TFIM)) or \(h \rightarrow 0\) (Longitudinal Field Ising Model (LFIM)), the Hamiltonian~\eqref{eq:TFIM} becomes integrable.
We study chaos and thermalization time scales in the vicinity of these limits in order to observe and quantify the possible differences between the two limits.

In classical systems near integrable limits, the way the actions are coupled by the integrability-breaking perturbations defines different classes of networks with different properties.
The system is defined as a long-range network if the connectivity (defined through an appropriately defined \emph{coupling range}) is extensive in the number of actions \(N\).
In a short-range network, the connectivity of the actions (i.e. the coupling range) is independent of the number of actions.

Inspired by the classical definition, we focus on the conserved quantities of the QIC in its integrable limits.
In the limit \(h = 0\), the spin chain becomes effectively decoupled in real space and the conserved quantities are local, with the simplest one being \(\sigma^{z}_{i}\).
Adding a small nonzero value of \(h\) will couple these conserved quantities in a local manner of a short-range network.

In the limit \(g = 0\), the spin chain is extensively connected while still being integrable.
This is reflected by the fact that the conserved quantities are non-local\footnote{Corresponding to the fermionic operators in the Jordan-Wigner representation of the system}.
Some of these operators correspond to simple symmetry operations.
For example, the spin-flip operation \(\sigma^z \rightarrow -\sigma^z\) for all spins leaves the Hamiltonian invariant.
The corresponding conserved quantity is \(\prod_{i = 1}^{N}\sigma^{x}_{i}\).
Other conserved quantities are similarly represented as extensive (non-local) combinations of the local spin matrices or as sums of local terms.
The support of such quantities grows with system size \footnote{Somewhat related to this is the delocalization in Fock space, which has been studied in Ref.~\onlinecite{bulchandani2022onset}}. 

Thus, by analogy with the classical definition~\cite{mithun2018weakly,mithun2019dynamical,danieli2019dynamical} we classify the quantum weakly non-integrable models by the character of coupling of conserved quantities in the integrable limit by the integrability breaking perturbation.
We consider the following two types of networks:
(i) \emph{Quantum short-range network} (SRN)---a conserved quantity in the integrable limit is coupled by the integrability breaking perturbation to a system size independent number of other conserved quantities, as observed from the operator dynamics defined by standard commutator relations. 
(ii) \emph{Quantum long-range network} (LRN)---a conserved quantity in the integrable limit is coupled by the integrability breaking perturbation to a number of other conserved quantities that scale with the system size.

In what follows, we probe the ergodicity (ETH) and operator growth timescales of the above operators near the two respective limits. 
We then compare the divergence of these two time scales upon approaching each of the two limits.

\sect{Krylov complexity}

There exists a large class of observables that quantify operator growth under Hamiltonian dynamics.
A common feature among most of these probes is the choice of a basis to expand the time-evolved operator. 
Once the basis is chosen, then appropriate \emph{expectation values} are defined and evaluated, which then serve to distinguish between chaotic and integrable systems. 

One such probe is Krylov complexity~\cite{parker2019A}.
The steps to evaluate Krylov complexity begin with generating a \emph{minimal basis}~\footnote{Minimal with respect to a cost function; which in this case is the Krylov complexity itself.}, which is called the Krylov basis. 
The Krylov complexity captures the average position of an operator in a minimal basis under the operator's unitary evolution in time with the Hamiltonian \(H\).
The construction of the Krylov basis relies on an appropriately chosen norm in the Hilbert space of operators.
We employ the infinite-temperature Hilbert-Schmidt norm for our analysis:
\begin{align}
    \label{eq:hsnorm}
    (A \vert B) = \frac{\text{Tr}\left(A^\dagger B \right)}{\mathcal{D}}.
\end{align}

Upon adopting the norm, one chooses an operator \(\mathcal{O}\) whose evolution is studied.
The unitary evolution of \(\mathcal{O}\) governed by the Hamiltonian \(H\) is defined as
\begin{align}
    \label{eq:opt}
    \mot = e^{i H t}\mop e^{-i H t} = e^{i \mathcal{L} t}\mop
\end{align}
where \(\mathcal{L}(*) \equiv [H, * ]\) is the Louivillian superoperator.
The elements \(\mop_{n}\) of the minimal (Krylov) basis corresponding to the operator \(\mop\) and Hamiltonian \(H\) are generated via the Lanczos algorithm~\cite{viswanath1994recursion, parker2019A}, as described in the Supplementary Material. 
The operator \(\mot\) has the following expansion in this basis
\begin{align}
    \mot = \sum_{n}i^{n}\psi_{n}(t)\mop_{n} .
\end{align}
The dynamical properties of \(\mot\) under the Hamiltonian \(H\) are encoded in the behavior of the Krylov wavefunctions \(\psi_n (t)\).
These properties allow to diagnose chaotic behavior in quantum many-body systems~\cite{parker2019A,bhattacharjee2022krylov,bhattacharjee2022probing,rabinovici2020operator,rabinovici2022krylov,rabinovici2022krylov2}.

It was argued in Ref.~\onlinecite{parker2019A} that the average position of the time-evolved operator
\begin{align}
    K(t) = \sum_{n}^{\mathcal{K}} n\vert \psi_{n}(t) \vert^2,
\end{align}
known as Krylov complexity, grows exponentially with \(t\) for chaotic dynamics: \(K(t) \sim e^{2 \alpha t}\).
The exponent \(\alpha\) captures the strength of the chaotic dynamics and is bounded from above by the Maladcena-Shenker-Stanford~\cite{maldacena2015A} bound.
One can then define a natural time-scale for a chaotic system as \(\alpha^{-1}\), which we denote as \(T_\lambda\) throughout this manuscript since it captures the growth of operators under the Hamiltonian \(H\).

\sect{Ergodization time}
Eigenstate thermalization hypothesis~\cite{srednicki1994chaos,deutsch2018eigenstate} provides a powerful tool to study and characterize thermalization in quantum mechanical systems.
It serves as a way to probe the chaotic or integrable dynamics of a system through the evolution of operators and states.
The essential statement of ETH can be encapsulated in the following equation
\begin{align}
    \label{eq:etheqn}
    \langle{\mot}\rangle\vert_{t\rightarrow \infty} = \omo + \frac{1}{\sqrt{\mathcal{D}}}R(t)
\end{align}
where the expectation value \(\langle{\cdots}\rangle\) of the time-evolved operator is taken in a typical state and the long time-averaged expectation value of the operator \(\omo\) can be evaluated analytically in the diagonal ensemble.
The function \(R(t)\) represents subleading fluctuations, suppressed by the Hilbert space dimension \(\mathcal{D}\).
The nature of the function \(R(t)\) has been studied extensively in different chaotic and integrable systems~\cite{zhang2022statistical,richter2020eigenstate,frizsch2021eigenstate,vidmar2021phenomenology}.
We use the fluctuations of \(R(t)\) to extract a timescale, that we refer to as the ergodization timescale \(T_{E}\).
The definition of this scale largely follows that in the classical case, discussed in Ref.~\onlinecite{malishava2022thermalization}.
We choose a random initial state \(\ket{\psi}\) as the typical state and track the evolution of \(\mot\).
As the function \(\langle \mot \rangle\) evolves it eventually starts to fluctuate around the mean value \(\overline{O}\), going above and below the mean with time.
This defines the passage times \(t_i\) of the function across the mean value: \(\langle\mathcal{O}(t_i)\rangle = \omo\)~\cite{mithun2018weakly,mithun2019dynamical,danieli2019dynamical,danieli2017intermittent,zhang2024thermalization,lando2023thermalization,malishava2022lyapunov,malishava2022thermalization}.
Time intervals between the two subsequent passages
\begin{gather}
    \label{eq:tau-def}
    \tau_i = t_{i + 1} - t_{i}
\end{gather}
are called excursion times, since they reflect the time spent by the expectation value \(\langle\mot\rangle\) away from its mean value.
We distinguish additionally positive, \(\tau_{i,+}\), and negative, \(\tau_{i,-}\) excursions times for excursions above and below the mean value \(\omo\).
For a given system and operator, the collected excursion times \(\tau_i\) obey some distribution.
We use the moments---mean and variance---of this distribution to extract an ergodization time scale, as defined later and discussed in the Supplementary Material.

\sect{Results}
We study numerically the dynamical properties close to integrability of the quantum short- and long-range networks respectively using the definitions and the methods outlined above.
The Hamiltonian is given by Eq.~\eqref{eq:TFIM} and we choose the following conserved quantities in the two integrable limits of the Hamiltonian as the time-evolved operators/observables whose dynamical properties we study.
{\color{black}
\begin{align}
    \label{eq:op1}
    \mop_{h \rightarrow 0} &= \sigma^{z}_{i} \\
    \label{eq:op2}
    \mop^{(k)}_{g \rightarrow 0} &= I^{(k)}\;\;\;\; k = 1,\dots,N
\end{align}
where $I^{(k)}$ denotes the set of conserved quantities for the Transverse field Ising model
\begin{align}
    I^{(k)} &= i J \sum_{j = 1}^{N}\left(S^{z y}_{j:j+k} - S^{y z}_{j:j+k}\right)\label{ik}
\end{align}
where we have the following shorthand
\begin{align}
    S^{\alpha \beta}_{j:j+l} = \sigma^{\alpha}_{j}\left(\prod_{n = 1}^{l - 1}\sigma^{x}_{j+n}\right)\sigma^{\beta}_{j + l}
\end{align}
The conserved quantity \(\mop^{(k)}_{h\rightarrow0}\) is comprised of sums of \(q\)-local quantities, which have support on \(q\) lattice sites.
For \(I^{(k)}\), we have \(q = k + 1\).
At the respective integrable limits, the operators are conserved, their corresponding Lyapunov exponents are \(0\) and their Lyapunov times are defined through the Krylov complexity diverge.
Similar to the classical case, we are interested in quantifying the divergence of the Lyapunov and ergodization times (from the two probes) upon approaching the integrable limits.
For the following discussion, we will present the timescales obtained by averaging over the \(N\) operators \(\mop^{(k)}_{g \rightarrow 0}\) in the LRN case.
The individual timescales are presented in the Supplementary.
For the SRN case, averaging over the \(N\) possible \(\sigma^{z}_{i}\) gives quantitatively similar timescales as that of an individual \(\sigma^{z}_{i}\).
This is discussed in the Supplementary.}

\begin{figure}
    \includegraphics[width=0.95\columnwidth]{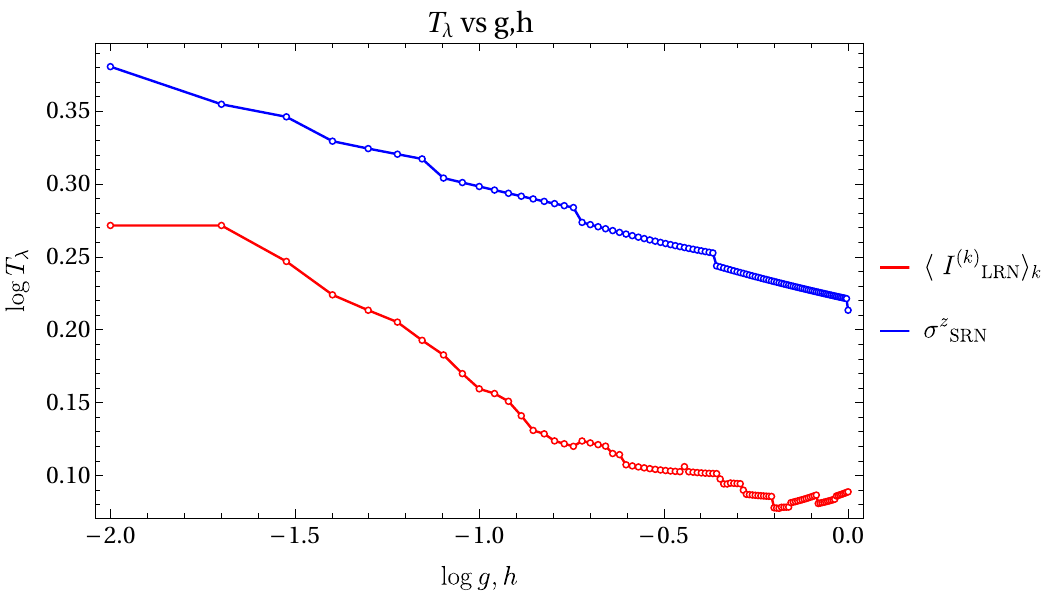}
    \caption{
        The Lyapunov time \(T_\lambda\) for \(N=8\) spins extracted from the linear growth of the Lanczos coefficients of time evolved operator \(\mot\) in the Krylov basis as a function of \(h\) for SRN and \(g\) for LRN (averaged) respectively.
        In both limits \(T_\lambda\) shows a clear increase upon approach to the integrable limit.
    }
    \label{fig:lrnsnrTL}
\end{figure}

Figure~\ref{fig:lrnsnrTL} shows the Lyapunov times \(T_\lambda\) extracted from the linear growth of the Lanczos coefficients in the Krylov basis of the operators~(\ref{eq:op1}-\ref{eq:op2}) and plotted in the log-log scale as functions of the integrability breaking parameters \(g\) (LRN) or \(h\) (SRN).
Our LRN data for \(T_\lambda\) are in semiquantitative agreement with similar computations in Ref.~\onlinecite{parker2019A}, although the Hamiltonian parameters do not completely match, and different operators were used.

In both LRN and SRN cases, the observed behavior of \(T_\lambda\) is fitted with a (weak) power-law divergence for small values of \(g\) or \(h\).
One may consider other fitting attempts for different model families in Ref.~\onlinecite{parker2019A} which involve logarithmic fits.
The differences are small in the considered parameter range, and all that matters for our purpose here is to use one and the same fitting procedure for all measured time scales.
The exponents are extracted for a system of \(N=8\) spins via a linear fit of the data on the log-log scale
\begin{align}
    \log T_{\lambda, SRN} &= -0.083*\log(h) + 0.218 \label{tlsrn} \\
    \log T_{\lambda,LRN} &= -0.155*\log(g) + 0.010. \label{tllrn}
\end{align}

To study the thermalization properties, we collect the statistics of the excursion times \(\tau_{\pm}\)~\eqref{eq:tau-def} of the expectation value of the respective operators~(\ref{eq:op1}-\ref{eq:op2}) (denoted by the subscript \(\sigma^{z}_{\text{SRN}}\) and \(\langle I^{(k)}_{\text{LRN}} \rangle_{k}\) in the numerical plots) in a random state (different choices of the state yielded similar results).
We average the timescales over the \(N\) conserved quantities \(I^{(k)}\) for the LRN case.
The results for individual quantities \(I^{(k)}\) (\ref{ik}) are discussed in the Supplementary.
We study the mean \(\mu\) and variance \(\sigma^2\) of the positive and negative excursion times~\eqref{eq:tau-def}. 
To define an appropriate ergodization time \(T_{E}\), we study the relative behavior of mean \(\mu\) and variance \(\sigma^2\), with the integrability breaking parameter, by collecting \(10^4\) excursions.
Increasing this number does not change the moments significantly.
The following behaviour of the moments is observed, which follows rather closely the classical weakly nonintegrable systems:
\updated{We observe that the values of \(\sigma\) are exponentially larger than \(\mu\) close to the integrable limit for both SRN and LRN, suggesting that the typical timescale of fluctuations is dominated by the distribution tail rather than its mean.
A natural ergodization timescale is then defined as \(T_{E} = \sigma^2/\mu\)~\cite{mithun2018weakly,mithun2019dynamical,danieli2019dynamical,danieli2017intermittent,zhang2024thermalization,lando2023thermalization,malishava2022lyapunov,malishava2022thermalization}.}

\begin{figure}
    \centering
    \includegraphics[width=0.95\columnwidth]{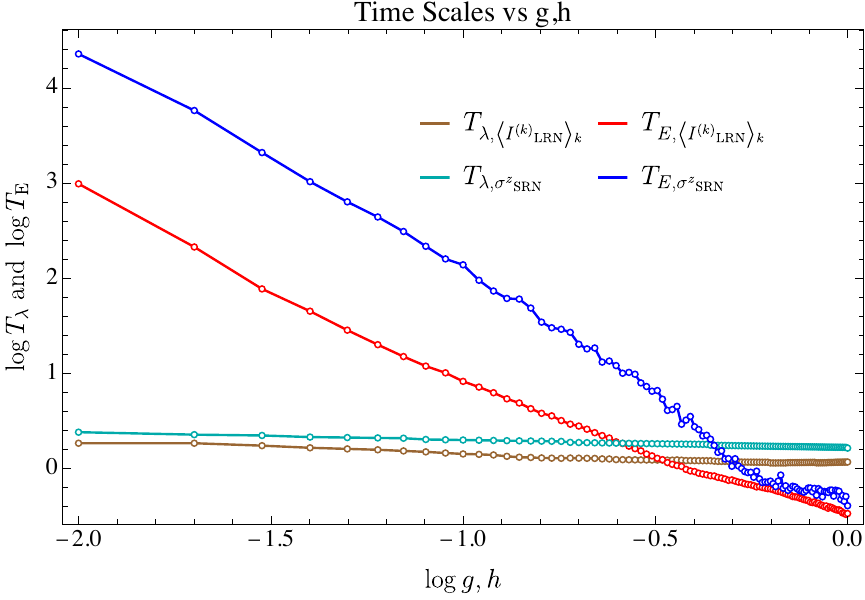}
    \caption{
        Comparison of the ergodization time \(T_{E}\) and Lyapunov time \(T_\lambda\) for the SRN (near-Ising limit) and LRN (near-Free limit).
    }
    \label{fig:comp2}
\end{figure}

In Fig.~\ref{fig:comp2} we compare the ergodization times \(T_{E,\pm}\) for the SRN and LRN based on positive/negative excursion times with the Lyapunov time \(T_\lambda\) obtained via the Krylov method.
Similarly to the Lyapunov time \(T_\lambda\) the ergodization time \(T_E\) also shows a power-law divergence with the decrease of the integrability breaking parameter.
The linear fits of the data in the log-log scale are
\begin{align}
    \log T_{E, SRN} &= -2.263*\log(h) -0.137 \label{tesrn}\\
    \log T_{E, LRN} &= -2.073*\log(g) - 1.150 \label{telrn}
\end{align}

Ergodization times extracted from positive and negative excursion times \(\tau_\pm\) show similar scaling close to integrability.
Here we discuss the result obtained from \(\tau_{+}\).

\updated{Our findings indicate that in the two integrable limits, the timescales \(T_\lambda\) and \(T_E\) diverge with exponents that differ by at least an order of magnitude.
This follows from comparing~\eqref{tlsrn} with~\eqref{tesrn} and~\eqref{tllrn} with~\eqref{telrn}. 
Therefore, the timescale associated with ETH diverges \emph{exponentially} faster than that associated with operator growth.
For the two network classes, the Lyapunov and Ergodization timescales respectively diverge with comparable exponents.
This suggests a universality in the mechanism of integrability breaking in quantum many-body spin systems. 
The short-range network, characterized by local conserved quantities in the integrable limit demonstrates a slowing-down of thermalization at a rate which appears to be similar to the long-range network which is characterized by non-local conserved quantities (which appear as sums of local terms) in the integrable limit.
However, such a conclusion can be partially deceiving, since we vary different parameters \(g\) and \(h\).}

\begin{figure}
    \centering
    \includegraphics[width=0.95\columnwidth]{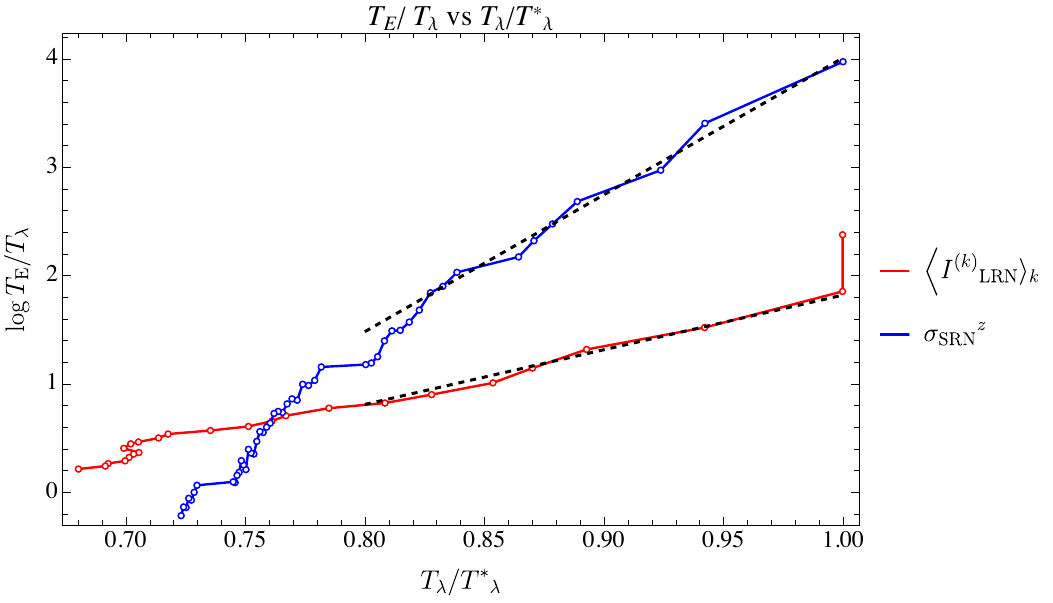}
    \caption{
        \(T_{E}/T_\lambda\) for LRN and SRN, plotted with respect to \(T_{\lambda}/T^{*}_{\lambda}\). The dashed lines represent the linear fit in the log-linear scale. The slopes correspond to $\alpha = 6.16$ for LRN (in red) and $\alpha = 12.63$ for SRN (in blue).  
    }
    \label{fig:lnrsnrfull}
\end{figure}
\updated{In order to properly compare the time scales from both network regimes, we replot them in units of the corresponding largest Lyapunov times. 
This is done in Fig.~\ref{fig:lnrsnrfull}, which is our central result and shows the ratio \(T_{E,\pm}/T_\lambda\) as a function of \(T_{\lambda}/T^{*}_{\lambda}\).
Now we put the time scale analysis of both classes on a similar footing \footnote{In particular, other forms of comparison have the underlying assumption that the integrability-breaking parameters \(g\) and \(h\) can be treated similarly.
Comparison of \(T_{E,\pm}/T_\lambda\) with \(T_{\lambda}/T^{*}_{\lambda}\) frees us of this assumption.}.
Here \(T^{*}_\lambda\) is the maximum value of \(T_\lambda\) for each (SRN, LRN) of the networks and is required for effective comparison of the two networks, since the range of \(T_\lambda\) observed in Fig.~\ref{fig:lrnsnrTL} is different for the SRN and LRN networks.
It is evident from Fig.~\ref{fig:lnrsnrfull} that for the two network classes, \(T_{E,\pm}/T_{\lambda}\) scales in qualitatively different ways as \(T_\lambda^{\alpha}\) with \(\alpha \gg 1\), especially near \(T_\lambda/T^{*}_\lambda = 1\).
We find a rather weak scaling \(\alpha=6.16\) for the LRN, as compared to a much stronger scaling \(\alpha=12.63\) for the SRN.
}

\sect{Conclusions}
In this manuscript, we \updated{investigated} the universality classes of thermalization of classical weakly non-integrable systems in the case of weakly non-integrable \emph{quantum} many-body spin systems.
The classes are defined by the two timescales quantifying thermalization:
one timescale comes from the Krylov complexity of operator growth.
In the non-integrable regime, the K-complexity grows as \(\exp(t/T_\lambda)\), defining a Lyapunov timescale.

Another timescale, the ergodization time \(T_E\), is inspired by ETH and is defined through the statistics of the time intervals between consecutive crossings of the expectation of time-evolved operator \(\langle \mot \rangle\) around its mean value \(\omo\).
\(T_E\) is then defined through the appropriate moments of the intervals.

For \(\mop\) conserved in the integrable limit, both timescales are expected to diverge.

The short-range network (SRN) is defined by the locality of interaction between the conserved quantities (in the integrable limit) as integrability is weakly broken.
We find that the distribution of the crossing intervals is fat-tailed in this case and therefore \(T_{E}\) is defined as the ratio of the variance and the mean.

Conversely, the long-range network (LRN) is defined by the non-locality of interaction between the conserved quantities upon weak integrability-breaking.
\updated{Similar to the classical observation~\cite{danieli2019dynamical}, we find that the two timescales respond in a qualitatively different way as compared to the SRN case, underscoring the difference of the integrability-breaking mechanisms}.

In the 1D spin-\(1/2\) chain we study, both timescales, \(T_\lambda\) and \(T_E\), diverge as power-laws with decreasing integrability breaking parameter.
We compare the exponents of both \(T_{E}\) and \(T_{\lambda}\) in the two network classes.
We find that the exponents of \(T_E\) and \(T_\lambda\) are comparable to each other for LRN and SRN.
\updated{However, this result uses varying different model parameters.
In order to be able to quantitatively compare the relative growth of the time scales for both regimes, we measure \(T_E\) in units of \(T_\lambda\) and plot the outcome as a function of \(T_\lambda\).
We then find that the LRN regime shows a relatively small rate at which thermalization (in the ETH sense) slows down upon approaching the integrable limit compared to operator spreading.
For the SRN case instead, the rate at which thermalization (in the ETH sense) slows down upon approaching the integrable limit compared to operator spreading, is much larger.
Therefore, in the SRN case it takes more and more time to thermalize as compared to the operator growth time scales.}

This implies that in the SRN regime, ETH-like thermalization slows down exponentially faster than operator spreading as integrability is approached.

We identify this drastic difference in the relative slow-down of thermalization and operator spreading as the \updated{universal} feature of short-range networks, as opposed to long-range networks.
This is \updated{similar} to the character of thermalization slowing down in classical systems \cite{malishava2022lyapunov,malishava2022thermalization} where a classical Lyapunov time is compared to the classical ergodization time, and \updated{the two network classes respond in very different manners}. 

Many open questions naturally emerge from this investigation.
One natural direction is testing this classification for other types of quantum systems.
Further, other probes might be able to distinguish and be sensitive to these two network classes.
In the classical case, the Lyapunov spectrum scaling close to integrability proved to be instrumental in classifying network classes~\cite{malishava2022lyapunov}.
It would interesting to study probes that do not have classical analogues, such as quantum entanglement~\cite{karthik2007entanglement}, in such phenomena.
Our investigation was restricted to finite system sizes, the scaling of the two timescales with system size (and therefore in the thermodynamic limit) is also worth investigating. 
There has also been a large body of work that has investigated the exact nature of the fluctuations of operator evolution (for example, \(R(t)\) in Eq.~\eqref{eq:etheqn}).
This suggests that a more concrete connection between ETH and ergodization time might exist.
Random matrix theory is also expected to play a crucial role in this characterization.
It would also be interesting to study the effective random matrix theory near integrability for the two network classes.

\sect{Acknowledgments}
The authors acknowledge Tilen Cadez, Barbara Dietz, Yeongjun Kim, Gabriel Lando, Aniket Patra, Sonu Verma and Weihua Zhang for relevant discussions.
We acknowledge the financial support from the Institute for Basic Science (IBS) in the Republic of Korea through the Project No. IBS-R024-D1.

\bibliography{ergodicity,general,mbl,local}

\begin{thebibliography}{90}%
\makeatletter
\providecommand \@ifxundefined [1]{%
 \@ifx{#1\undefined}
}%
\providecommand \@ifnum [1]{%
 \ifnum #1\expandafter \@firstoftwo
 \else \expandafter \@secondoftwo
 \fi
}%
\providecommand \@ifx [1]{%
 \ifx #1\expandafter \@firstoftwo
 \else \expandafter \@secondoftwo
 \fi
}%
\providecommand \natexlab [1]{#1}%
\providecommand \enquote  [1]{``#1''}%
\providecommand \bibnamefont  [1]{#1}%
\providecommand \bibfnamefont [1]{#1}%
\providecommand \citenamefont [1]{#1}%
\providecommand \href@noop [0]{\@secondoftwo}%
\providecommand \href [0]{\begingroup \@sanitize@url \@href}%
\providecommand \@href[1]{\@@startlink{#1}\@@href}%
\providecommand \@@href[1]{\endgroup#1\@@endlink}%
\providecommand \@sanitize@url [0]{\catcode `\\12\catcode `\$12\catcode
  `\&12\catcode `\#12\catcode `\^12\catcode `\_12\catcode `\%12\relax}%
\providecommand \@@startlink[1]{}%
\providecommand \@@endlink[0]{}%
\providecommand \url  [0]{\begingroup\@sanitize@url \@url }%
\providecommand \@url [1]{\endgroup\@href {#1}{\urlprefix }}%
\providecommand \urlprefix  [0]{URL }%
\providecommand \Eprint [0]{\href }%
\providecommand \doibase [0]{https://doi.org/}%
\providecommand \selectlanguage [0]{\@gobble}%
\providecommand \bibinfo  [0]{\@secondoftwo}%
\providecommand \bibfield  [0]{\@secondoftwo}%
\providecommand \translation [1]{[#1]}%
\providecommand \BibitemOpen [0]{}%
\providecommand \bibitemStop [0]{}%
\providecommand \bibitemNoStop [0]{.\EOS\space}%
\providecommand \EOS [0]{\spacefactor3000\relax}%
\providecommand \BibitemShut  [1]{\csname bibitem#1\endcsname}%
\let\auto@bib@innerbib\@empty
\bibitem [{\citenamefont {Bohigas}\ \emph {et~al.}(1984)\citenamefont
  {Bohigas}, \citenamefont {Giannoni},\ and\ \citenamefont
  {Schmit}}]{bohigas1984characterization}%
  \BibitemOpen
  \bibfield  {author} {\bibinfo {author} {\bibfnamefont {O.}~\bibnamefont
  {Bohigas}}, \bibinfo {author} {\bibfnamefont {M.~J.}\ \bibnamefont
  {Giannoni}},\ and\ \bibinfo {author} {\bibfnamefont {C.}~\bibnamefont
  {Schmit}},\ }\bibfield  {title} {\bibinfo {title} {Characterization of
  chaotic quantum spectra and universality of level fluctuation laws},\ }\href
  {https://doi.org/10.1103/PhysRevLett.52.1} {\bibfield  {journal} {\bibinfo
  {journal} {Phys. Rev. Lett.}\ }\textbf {\bibinfo {volume} {52}},\ \bibinfo
  {pages} {1} (\bibinfo {year} {1984})}\BibitemShut {NoStop}%
\bibitem [{\citenamefont {Berry}\ \emph {et~al.}(1977)\citenamefont {Berry},
  \citenamefont {Tabor},\ and\ \citenamefont {Ziman}}]{berry1977level}%
  \BibitemOpen
  \bibfield  {author} {\bibinfo {author} {\bibfnamefont {M.~V.}\ \bibnamefont
  {Berry}}, \bibinfo {author} {\bibfnamefont {M.}~\bibnamefont {Tabor}},\ and\
  \bibinfo {author} {\bibfnamefont {J.~M.}\ \bibnamefont {Ziman}},\ }\bibfield
  {title} {\bibinfo {title} {Level clustering in the regular spectrum},\ }\href
  {https://doi.org/10.1098/rspa.1977.0140} {\bibfield  {journal} {\bibinfo
  {journal} {Proceedings of the Royal Society of London. A. Mathematical and
  Physical Sciences}\ }\textbf {\bibinfo {volume} {356}},\ \bibinfo {pages}
  {375} (\bibinfo {year} {1977})},\ \Eprint
  {https://arxiv.org/abs/https://royalsocietypublishing.org/doi/pdf/10.1098/rspa.1977.0140}
  {https://royalsocietypublishing.org/doi/pdf/10.1098/rspa.1977.0140}
  \BibitemShut {NoStop}%
\bibitem [{\citenamefont {Zakrzewski}(2023)}]{zakrewski2023quantum}%
  \BibitemOpen
  \bibfield  {author} {\bibinfo {author} {\bibfnamefont {J.}~\bibnamefont
  {Zakrzewski}},\ }\bibfield  {title} {\bibinfo {title} {{Quantum Chaos and
  Level Dynamics}},\ }\href {https://doi.org/10.3390/e25030491} {\bibfield
  {journal} {\bibinfo  {journal} {Entropy}\ }\textbf {\bibinfo {volume} {25}},\
  \bibinfo {pages} {491} (\bibinfo {year} {2023})},\ \Eprint
  {https://arxiv.org/abs/2302.05934} {arXiv:2302.05934 [cond-mat.dis-nn]}
  \BibitemShut {NoStop}%
\bibitem [{\citenamefont {Haake}(2006)}]{haake2006quantum}%
  \BibitemOpen
  \bibfield  {author} {\bibinfo {author} {\bibfnamefont {F.}~\bibnamefont
  {Haake}},\ }\href@noop {} {\emph {\bibinfo {title} {Quantum Signatures of
  Chaos}}}\ (\bibinfo  {publisher} {Springer-Verlag},\ \bibinfo {address}
  {Berlin, Heidelberg},\ \bibinfo {year} {2006})\BibitemShut {NoStop}%
\bibitem [{\citenamefont {Br\'ezin}\ and\ \citenamefont
  {Hikami}(1997)}]{brezin1997spectral}%
  \BibitemOpen
  \bibfield  {author} {\bibinfo {author} {\bibfnamefont {E.}~\bibnamefont
  {Br\'ezin}}\ and\ \bibinfo {author} {\bibfnamefont {S.}~\bibnamefont
  {Hikami}},\ }\bibfield  {title} {\bibinfo {title} {Spectral form factor in a
  random matrix theory},\ }\href {https://doi.org/10.1103/PhysRevE.55.4067}
  {\bibfield  {journal} {\bibinfo  {journal} {Phys. Rev. E}\ }\textbf {\bibinfo
  {volume} {55}},\ \bibinfo {pages} {4067} (\bibinfo {year} {1997})},\ \Eprint
  {https://arxiv.org/abs/cond-mat/9608116} {arXiv:cond-mat/9608116 [cond-mat]}
  \BibitemShut {NoStop}%
\bibitem [{\citenamefont {Garc\'\i{}a-Garc\'\i{}a}\ and\ \citenamefont
  {Verbaarschot}(2016)}]{garcia2016spectral2}%
  \BibitemOpen
  \bibfield  {author} {\bibinfo {author} {\bibfnamefont {A.~M.}\ \bibnamefont
  {Garc\'\i{}a-Garc\'\i{}a}}\ and\ \bibinfo {author} {\bibfnamefont {J.~J.~M.}\
  \bibnamefont {Verbaarschot}},\ }\bibfield  {title} {\bibinfo {title}
  {{Spectral and thermodynamic properties of the Sachdev-Ye-Kitaev model}},\
  }\href {https://doi.org/10.1103/PhysRevD.94.126010} {\bibfield  {journal}
  {\bibinfo  {journal} {Phys. Rev. D}\ }\textbf {\bibinfo {volume} {94}},\
  \bibinfo {pages} {126010} (\bibinfo {year} {2016})},\ \Eprint
  {https://arxiv.org/abs/1610.03816} {arXiv:1610.03816 [hep-th]} \BibitemShut
  {NoStop}%
\bibitem [{\citenamefont {Garc\'\i{}a-Garc\'\i{}a}\ and\ \citenamefont
  {Verbaarschot}(2017)}]{garcia2017spectral}%
  \BibitemOpen
  \bibfield  {author} {\bibinfo {author} {\bibfnamefont {A.~M.}\ \bibnamefont
  {Garc\'\i{}a-Garc\'\i{}a}}\ and\ \bibinfo {author} {\bibfnamefont {J.~J.~M.}\
  \bibnamefont {Verbaarschot}},\ }\bibfield  {title} {\bibinfo {title}
  {{Analytical Spectral Density of the Sachdev-Ye-Kitaev Model at finite N}},\
  }\href {https://doi.org/10.1103/PhysRevD.96.066012} {\bibfield  {journal}
  {\bibinfo  {journal} {Phys. Rev. D}\ }\textbf {\bibinfo {volume} {96}},\
  \bibinfo {pages} {066012} (\bibinfo {year} {2017})},\ \Eprint
  {https://arxiv.org/abs/1701.06593} {arXiv:1701.06593 [hep-th]} \BibitemShut
  {NoStop}%
\bibitem [{\citenamefont {Khramtsov}\ and\ \citenamefont
  {Lanina}(2021)}]{khramtsov2021spectral}%
  \BibitemOpen
  \bibfield  {author} {\bibinfo {author} {\bibfnamefont {M.}~\bibnamefont
  {Khramtsov}}\ and\ \bibinfo {author} {\bibfnamefont {E.}~\bibnamefont
  {Lanina}},\ }\bibfield  {title} {\bibinfo {title} {{Spectral form factor in
  the double-scaled SYK model}},\ }\href
  {https://doi.org/10.1007/JHEP03(2021)031} {\bibfield  {journal} {\bibinfo
  {journal} {JHEP}\ }\textbf {\bibinfo {volume} {03}},\ \bibinfo {pages}
  {031}},\ \Eprint {https://arxiv.org/abs/2011.01906} {arXiv:2011.01906
  [hep-th]} \BibitemShut {NoStop}%
\bibitem [{\citenamefont {Forrester}(2010)}]{forrester2010log}%
  \BibitemOpen
  \bibfield  {author} {\bibinfo {author} {\bibfnamefont {P.~J.}\ \bibnamefont
  {Forrester}},\ }\href {https://doi.org/doi:10.1515/9781400835416} {\emph
  {\bibinfo {title} {Log-Gases and Random Matrices (LMS-34)}}}\ (\bibinfo
  {publisher} {Princeton University Press},\ \bibinfo {address} {Princeton},\
  \bibinfo {year} {2010})\BibitemShut {NoStop}%
\bibitem [{\citenamefont {Mehta}(2004)}]{mehta2004random}%
  \BibitemOpen
  \bibfield  {author} {\bibinfo {author} {\bibfnamefont {M.~L.}\ \bibnamefont
  {Mehta}},\ }\href@noop {} {\emph {\bibinfo {title} {Random matrices}}},\
  \bibinfo {edition} {3rd}\ ed.,\ Pure and applied mathematics: v. 142\
  (\bibinfo  {publisher} {Elsevier/Academic Press},\ \bibinfo {address}
  {Amsterdam},\ \bibinfo {year} {2004})\BibitemShut {NoStop}%
\bibitem [{\citenamefont {Stanford}(2016)}]{stanford2015many}%
  \BibitemOpen
  \bibfield  {author} {\bibinfo {author} {\bibfnamefont {D.}~\bibnamefont
  {Stanford}},\ }\bibfield  {title} {\bibinfo {title} {{Many-body chaos at weak
  coupling}},\ }\href {https://doi.org/10.1007/JHEP10(2016)009} {\bibfield
  {journal} {\bibinfo  {journal} {JHEP}\ }\textbf {\bibinfo {volume} {10}},\
  \bibinfo {pages} {009}},\ \Eprint {https://arxiv.org/abs/1512.07687}
  {arXiv:1512.07687 [hep-th]} \BibitemShut {NoStop}%
\bibitem [{\citenamefont {Caputa}\ \emph {et~al.}(2016)\citenamefont {Caputa},
  \citenamefont {Numasawa},\ and\ \citenamefont
  {Veliz-Osorio}}]{caputa2016out}%
  \BibitemOpen
  \bibfield  {author} {\bibinfo {author} {\bibfnamefont {P.}~\bibnamefont
  {Caputa}}, \bibinfo {author} {\bibfnamefont {T.}~\bibnamefont {Numasawa}},\
  and\ \bibinfo {author} {\bibfnamefont {A.}~\bibnamefont {Veliz-Osorio}},\
  }\bibfield  {title} {\bibinfo {title} {{Out-of-time-ordered correlators and
  purity in rational conformal field theories}},\ }\href
  {https://doi.org/10.1093/ptep/ptw157} {\bibfield  {journal} {\bibinfo
  {journal} {PTEP}\ }\textbf {\bibinfo {volume} {2016}},\ \bibinfo {pages}
  {113B06} (\bibinfo {year} {2016})},\ \Eprint
  {https://arxiv.org/abs/1602.06542} {arXiv:1602.06542 [hep-th]} \BibitemShut
  {NoStop}%
\bibitem [{\citenamefont {Roberts}\ and\ \citenamefont
  {Yoshida}(2017)}]{roberts2016chaos}%
  \BibitemOpen
  \bibfield  {author} {\bibinfo {author} {\bibfnamefont {D.~A.}\ \bibnamefont
  {Roberts}}\ and\ \bibinfo {author} {\bibfnamefont {B.}~\bibnamefont
  {Yoshida}},\ }\bibfield  {title} {\bibinfo {title} {{Chaos and complexity by
  design}},\ }\href {https://doi.org/10.1007/JHEP04(2017)121} {\bibfield
  {journal} {\bibinfo  {journal} {JHEP}\ }\textbf {\bibinfo {volume} {04}},\
  \bibinfo {pages} {121}},\ \Eprint {https://arxiv.org/abs/1610.04903}
  {arXiv:1610.04903 [quant-ph]} \BibitemShut {NoStop}%
\bibitem [{\citenamefont {Shen}\ \emph {et~al.}(2017)\citenamefont {Shen},
  \citenamefont {Zhang}, \citenamefont {Fan},\ and\ \citenamefont
  {Zhai}}]{shen2016out}%
  \BibitemOpen
  \bibfield  {author} {\bibinfo {author} {\bibfnamefont {H.}~\bibnamefont
  {Shen}}, \bibinfo {author} {\bibfnamefont {P.}~\bibnamefont {Zhang}},
  \bibinfo {author} {\bibfnamefont {R.}~\bibnamefont {Fan}},\ and\ \bibinfo
  {author} {\bibfnamefont {H.}~\bibnamefont {Zhai}},\ }\bibfield  {title}
  {\bibinfo {title} {{Out-of-Time-Order Correlation at a Quantum Phase
  Transition}},\ }\href {https://doi.org/10.1103/PhysRevB.96.054503} {\bibfield
   {journal} {\bibinfo  {journal} {Phys. Rev. B}\ }\textbf {\bibinfo {volume}
  {96}},\ \bibinfo {pages} {054503} (\bibinfo {year} {2017})},\ \Eprint
  {https://arxiv.org/abs/1608.02438} {arXiv:1608.02438 [cond-mat.quant-gas]}
  \BibitemShut {NoStop}%
\bibitem [{\citenamefont {Cotler}\ \emph {et~al.}(2017)\citenamefont {Cotler},
  \citenamefont {Hunter-Jones}, \citenamefont {Liu},\ and\ \citenamefont
  {Yoshida}}]{cotler2017chaos}%
  \BibitemOpen
  \bibfield  {author} {\bibinfo {author} {\bibfnamefont {J.}~\bibnamefont
  {Cotler}}, \bibinfo {author} {\bibfnamefont {N.}~\bibnamefont
  {Hunter-Jones}}, \bibinfo {author} {\bibfnamefont {J.}~\bibnamefont {Liu}},\
  and\ \bibinfo {author} {\bibfnamefont {B.}~\bibnamefont {Yoshida}},\
  }\bibfield  {title} {\bibinfo {title} {{Chaos, Complexity, and Random
  Matrices}},\ }\href {https://doi.org/10.1007/JHEP11(2017)048} {\bibfield
  {journal} {\bibinfo  {journal} {JHEP}\ }\textbf {\bibinfo {volume} {11}},\
  \bibinfo {pages} {048}},\ \Eprint {https://arxiv.org/abs/1706.05400}
  {arXiv:1706.05400 [hep-th]} \BibitemShut {NoStop}%
\bibitem [{\citenamefont {Cotler}\ \emph {et~al.}(2018)\citenamefont {Cotler},
  \citenamefont {Ding},\ and\ \citenamefont {Penington}}]{cotler2017out}%
  \BibitemOpen
  \bibfield  {author} {\bibinfo {author} {\bibfnamefont {J.~S.}\ \bibnamefont
  {Cotler}}, \bibinfo {author} {\bibfnamefont {D.}~\bibnamefont {Ding}},\ and\
  \bibinfo {author} {\bibfnamefont {G.~R.}\ \bibnamefont {Penington}},\
  }\bibfield  {title} {\bibinfo {title} {{Out-of-time-order Operators and the
  Butterfly Effect}},\ }\href {https://doi.org/10.1016/j.aop.2018.07.020}
  {\bibfield  {journal} {\bibinfo  {journal} {Annals Phys.}\ }\textbf {\bibinfo
  {volume} {396}},\ \bibinfo {pages} {318} (\bibinfo {year} {2018})},\ \Eprint
  {https://arxiv.org/abs/1704.02979} {arXiv:1704.02979 [quant-ph]} \BibitemShut
  {NoStop}%
\bibitem [{\citenamefont {Hashimoto}\ \emph {et~al.}(2017)\citenamefont
  {Hashimoto}, \citenamefont {Murata},\ and\ \citenamefont
  {Yoshii}}]{hashimoto2017out}%
  \BibitemOpen
  \bibfield  {author} {\bibinfo {author} {\bibfnamefont {K.}~\bibnamefont
  {Hashimoto}}, \bibinfo {author} {\bibfnamefont {K.}~\bibnamefont {Murata}},\
  and\ \bibinfo {author} {\bibfnamefont {R.}~\bibnamefont {Yoshii}},\
  }\bibfield  {title} {\bibinfo {title} {{Out-of-time-order correlators in
  quantum mechanics}},\ }\href {https://doi.org/10.1007/JHEP10(2017)138}
  {\bibfield  {journal} {\bibinfo  {journal} {JHEP}\ }\textbf {\bibinfo
  {volume} {10}},\ \bibinfo {pages} {138}},\ \Eprint
  {https://arxiv.org/abs/1703.09435} {arXiv:1703.09435 [hep-th]} \BibitemShut
  {NoStop}%
\bibitem [{\citenamefont {Fan}\ \emph {et~al.}(2017)\citenamefont {Fan},
  \citenamefont {Zhang}, \citenamefont {Shen},\ and\ \citenamefont
  {Zhai}}]{fan2017out}%
  \BibitemOpen
  \bibfield  {author} {\bibinfo {author} {\bibfnamefont {R.}~\bibnamefont
  {Fan}}, \bibinfo {author} {\bibfnamefont {P.}~\bibnamefont {Zhang}}, \bibinfo
  {author} {\bibfnamefont {H.}~\bibnamefont {Shen}},\ and\ \bibinfo {author}
  {\bibfnamefont {H.}~\bibnamefont {Zhai}},\ }\bibfield  {title} {\bibinfo
  {title} {Out-of-time-order correlation for many-body localization},\ }\href
  {https://doi.org/10.1016/j.scib.2017.04.011} {\bibfield  {journal} {\bibinfo
  {journal} {Science Bulletin}\ }\textbf {\bibinfo {volume} {62}},\ \bibinfo
  {pages} {707–711} (\bibinfo {year} {2017})},\ \Eprint
  {https://arxiv.org/abs/1608.01914} {arXiv:1608.01914 [cond-mat.quant-gas]}
  \BibitemShut {NoStop}%
\bibitem [{\citenamefont {Sun}\ \emph {et~al.}(2020)\citenamefont {Sun},
  \citenamefont {Cai}, \citenamefont {Tang}, \citenamefont {Hu},\ and\
  \citenamefont {Fan}}]{sun2018out}%
  \BibitemOpen
  \bibfield  {author} {\bibinfo {author} {\bibfnamefont {Z.-H.}\ \bibnamefont
  {Sun}}, \bibinfo {author} {\bibfnamefont {J.-Q.}\ \bibnamefont {Cai}},
  \bibinfo {author} {\bibfnamefont {Q.-C.}\ \bibnamefont {Tang}}, \bibinfo
  {author} {\bibfnamefont {Y.}~\bibnamefont {Hu}},\ and\ \bibinfo {author}
  {\bibfnamefont {H.}~\bibnamefont {Fan}},\ }\bibfield  {title} {\bibinfo
  {title} {Out-of-time-order correlators and quantum phase transitions in the
  rabi and dicke models},\ }\href
  {https://doi.org/https://doi.org/10.1002/andp.201900270} {\bibfield
  {journal} {\bibinfo  {journal} {Annalen der Physik}\ }\textbf {\bibinfo
  {volume} {532}},\ \bibinfo {pages} {1900270} (\bibinfo {year} {2020})},\
  \Eprint {https://arxiv.org/abs/1811.11191} {arXiv:1811.11191 [quant-ph]}
  \BibitemShut {NoStop}%
\bibitem [{\citenamefont {Tsuji}\ \emph {et~al.}(2018)\citenamefont {Tsuji},
  \citenamefont {Shitara},\ and\ \citenamefont {Ueda}}]{tsuji2018out}%
  \BibitemOpen
  \bibfield  {author} {\bibinfo {author} {\bibfnamefont {N.}~\bibnamefont
  {Tsuji}}, \bibinfo {author} {\bibfnamefont {T.}~\bibnamefont {Shitara}},\
  and\ \bibinfo {author} {\bibfnamefont {M.}~\bibnamefont {Ueda}},\ }\bibfield
  {title} {\bibinfo {title} {Out-of-time-order fluctuation-dissipation
  theorem},\ }\href {https://doi.org/10.1103/PhysRevE.97.012101} {\bibfield
  {journal} {\bibinfo  {journal} {Phys. Rev. E}\ }\textbf {\bibinfo {volume}
  {97}},\ \bibinfo {pages} {012101} (\bibinfo {year} {2018})}\BibitemShut
  {NoStop}%
\bibitem [{\citenamefont {de~Mello~Koch}\ \emph {et~al.}(2019)\citenamefont
  {de~Mello~Koch}, \citenamefont {Huang}, \citenamefont {Ma},\ and\
  \citenamefont {Van~Zyl}}]{deMelloKoch2019spectral}%
  \BibitemOpen
  \bibfield  {author} {\bibinfo {author} {\bibfnamefont {R.}~\bibnamefont
  {de~Mello~Koch}}, \bibinfo {author} {\bibfnamefont {J.-H.}\ \bibnamefont
  {Huang}}, \bibinfo {author} {\bibfnamefont {C.-T.}\ \bibnamefont {Ma}},\ and\
  \bibinfo {author} {\bibfnamefont {H.~J.~R.}\ \bibnamefont {Van~Zyl}},\
  }\bibfield  {title} {\bibinfo {title} {{Spectral Form Factor as an OTOC
  Averaged over the Heisenberg Group}},\ }\href
  {https://doi.org/10.1016/j.physletb.2019.06.025} {\bibfield  {journal}
  {\bibinfo  {journal} {Phys. Lett. B}\ }\textbf {\bibinfo {volume} {795}},\
  \bibinfo {pages} {183} (\bibinfo {year} {2019})},\ \Eprint
  {https://arxiv.org/abs/1905.10981} {arXiv:1905.10981 [hep-th]} \BibitemShut
  {NoStop}%
\bibitem [{\citenamefont {Akutagawa}\ \emph {et~al.}(2020)\citenamefont
  {Akutagawa}, \citenamefont {Hashimoto}, \citenamefont {Sasaki},\ and\
  \citenamefont {Watanabe}}]{akutagawa2020out}%
  \BibitemOpen
  \bibfield  {author} {\bibinfo {author} {\bibfnamefont {T.}~\bibnamefont
  {Akutagawa}}, \bibinfo {author} {\bibfnamefont {K.}~\bibnamefont
  {Hashimoto}}, \bibinfo {author} {\bibfnamefont {T.}~\bibnamefont {Sasaki}},\
  and\ \bibinfo {author} {\bibfnamefont {R.}~\bibnamefont {Watanabe}},\
  }\bibfield  {title} {\bibinfo {title} {{Out-of-time-order correlator in
  coupled harmonic oscillators}},\ }\href
  {https://doi.org/10.1007/JHEP08(2020)013} {\bibfield  {journal} {\bibinfo
  {journal} {JHEP}\ }\textbf {\bibinfo {volume} {08}},\ \bibinfo {pages}
  {013}},\ \Eprint {https://arxiv.org/abs/2004.04381} {arXiv:2004.04381
  [hep-th]} \BibitemShut {NoStop}%
\bibitem [{\citenamefont {Hu}\ and\ \citenamefont {Wan}(2021)}]{hu2020out}%
  \BibitemOpen
  \bibfield  {author} {\bibinfo {author} {\bibfnamefont {J.}~\bibnamefont
  {Hu}}\ and\ \bibinfo {author} {\bibfnamefont {S.}~\bibnamefont {Wan}},\
  }\bibfield  {title} {\bibinfo {title} {Out-of-time-ordered correlation in the
  anisotropic dicke model},\ }\href {https://doi.org/10.1088/1572-9494/ac256d}
  {\bibfield  {journal} {\bibinfo  {journal} {Communications in Theoretical
  Physics}\ }\textbf {\bibinfo {volume} {73}},\ \bibinfo {pages} {125703}
  (\bibinfo {year} {2021})},\ \Eprint {https://arxiv.org/abs/2008.10253}
  {arXiv:2008.10253 [cond-mat.quant-gas]} \BibitemShut {NoStop}%
\bibitem [{\citenamefont {Haferkamp}\ \emph {et~al.}(2022)\citenamefont
  {Haferkamp}, \citenamefont {Faist}, \citenamefont {Kothakonda}, \citenamefont
  {Eisert},\ and\ \citenamefont {Halpern}}]{haferkamp2021linear}%
  \BibitemOpen
  \bibfield  {author} {\bibinfo {author} {\bibfnamefont {J.}~\bibnamefont
  {Haferkamp}}, \bibinfo {author} {\bibfnamefont {P.}~\bibnamefont {Faist}},
  \bibinfo {author} {\bibfnamefont {N.~B.~T.}\ \bibnamefont {Kothakonda}},
  \bibinfo {author} {\bibfnamefont {J.}~\bibnamefont {Eisert}},\ and\ \bibinfo
  {author} {\bibfnamefont {N.~Y.}\ \bibnamefont {Halpern}},\ }\bibfield
  {title} {\bibinfo {title} {{Linear growth of quantum circuit complexity}},\
  }\href {https://doi.org/10.1038/s41567-022-01539-6} {\bibfield  {journal}
  {\bibinfo  {journal} {Nature Phys.}\ }\textbf {\bibinfo {volume} {18}},\
  \bibinfo {pages} {528} (\bibinfo {year} {2022})},\ \Eprint
  {https://arxiv.org/abs/2106.05305} {arXiv:2106.05305 [quant-ph]} \BibitemShut
  {NoStop}%
\bibitem [{\citenamefont {Mag\'an}(2018)}]{magan2018black}%
  \BibitemOpen
  \bibfield  {author} {\bibinfo {author} {\bibfnamefont {J.~M.}\ \bibnamefont
  {Mag\'an}},\ }\bibfield  {title} {\bibinfo {title} {{Black holes, complexity
  and quantum chaos}},\ }\href {https://doi.org/10.1007/JHEP09(2018)043}
  {\bibfield  {journal} {\bibinfo  {journal} {JHEP}\ }\textbf {\bibinfo
  {volume} {09}},\ \bibinfo {pages} {043}},\ \Eprint
  {https://arxiv.org/abs/1805.05839} {arXiv:1805.05839 [hep-th]} \BibitemShut
  {NoStop}%
\bibitem [{\citenamefont {Chapman}\ and\ \citenamefont
  {Policastro}(2022)}]{chapman2021quantum}%
  \BibitemOpen
  \bibfield  {author} {\bibinfo {author} {\bibfnamefont {S.}~\bibnamefont
  {Chapman}}\ and\ \bibinfo {author} {\bibfnamefont {G.}~\bibnamefont
  {Policastro}},\ }\bibfield  {title} {\bibinfo {title} {{Quantum computational
  complexity from quantum information to black holes and back}},\ }\href
  {https://doi.org/10.1140/epjc/s10052-022-10037-1} {\bibfield  {journal}
  {\bibinfo  {journal} {Eur. Phys. J. C}\ }\textbf {\bibinfo {volume} {82}},\
  \bibinfo {pages} {128} (\bibinfo {year} {2022})},\ \Eprint
  {https://arxiv.org/abs/2110.14672} {arXiv:2110.14672 [hep-th]} \BibitemShut
  {NoStop}%
\bibitem [{\citenamefont {Roca-Jerat}\ \emph {et~al.}(2023)\citenamefont
  {Roca-Jerat}, \citenamefont {Sancho-Lorente}, \citenamefont {Román-Roche},\
  and\ \citenamefont {Zueco}}]{rocajerat2023circuit}%
  \BibitemOpen
  \bibfield  {author} {\bibinfo {author} {\bibfnamefont {S.}~\bibnamefont
  {Roca-Jerat}}, \bibinfo {author} {\bibfnamefont {T.}~\bibnamefont
  {Sancho-Lorente}}, \bibinfo {author} {\bibfnamefont {J.}~\bibnamefont
  {Román-Roche}},\ and\ \bibinfo {author} {\bibfnamefont {D.}~\bibnamefont
  {Zueco}},\ }\bibfield  {title} {\bibinfo {title} {{Circuit Complexity through
  phase transitions: Consequences in quantum state preparation}},\ }\href
  {https://doi.org/10.21468/SciPostPhys.15.5.186} {\bibfield  {journal}
  {\bibinfo  {journal} {SciPost Phys.}\ }\textbf {\bibinfo {volume} {15}},\
  \bibinfo {pages} {186} (\bibinfo {year} {2023})}\BibitemShut {NoStop}%
\bibitem [{\citenamefont {Zhang}\ and\ \citenamefont
  {Yu}(2023)}]{zhang2022dynamical}%
  \BibitemOpen
  \bibfield  {author} {\bibinfo {author} {\bibfnamefont {P.}~\bibnamefont
  {Zhang}}\ and\ \bibinfo {author} {\bibfnamefont {Z.}~\bibnamefont {Yu}},\
  }\bibfield  {title} {\bibinfo {title} {Dynamical transition of operator size
  growth in quantum systems embedded in an environment},\ }\href
  {https://doi.org/10.1103/PhysRevLett.130.250401} {\bibfield  {journal}
  {\bibinfo  {journal} {Phys. Rev. Lett.}\ }\textbf {\bibinfo {volume} {130}},\
  \bibinfo {pages} {250401} (\bibinfo {year} {2023})},\ \Eprint
  {https://arxiv.org/abs/2211.03535} {arXiv:2211.03535 [quant-ph]} \BibitemShut
  {NoStop}%
\bibitem [{\citenamefont {Agarwal}\ and\ \citenamefont
  {Xu}(2020)}]{agarwal2020emergent}%
  \BibitemOpen
  \bibfield  {author} {\bibinfo {author} {\bibfnamefont {L.}~\bibnamefont
  {Agarwal}}\ and\ \bibinfo {author} {\bibfnamefont {S.}~\bibnamefont {Xu}},\
  }\bibfield  {title} {\bibinfo {title} {{Emergent symmetry in Brownian SYK
  models and charge dependent scrambling}},\ }\href
  {https://doi.org/10.1007/JHEP02(2022)045} {\bibfield  {journal} {\bibinfo
  {journal} {JHEP}\ }\textbf {\bibinfo {volume} {22}},\ \bibinfo {pages}
  {045}},\ \Eprint {https://arxiv.org/abs/2108.05810} {arXiv:2108.05810
  [cond-mat.str-el]} \BibitemShut {NoStop}%
\bibitem [{\citenamefont {Roberts}\ \emph {et~al.}(2018)\citenamefont
  {Roberts}, \citenamefont {Stanford},\ and\ \citenamefont
  {Streicher}}]{roberts2018operator}%
  \BibitemOpen
  \bibfield  {author} {\bibinfo {author} {\bibfnamefont {D.~A.}\ \bibnamefont
  {Roberts}}, \bibinfo {author} {\bibfnamefont {D.}~\bibnamefont {Stanford}},\
  and\ \bibinfo {author} {\bibfnamefont {A.}~\bibnamefont {Streicher}},\
  }\bibfield  {title} {\bibinfo {title} {Operator growth in the syk model},\
  }\href {https://api.semanticscholar.org/CorpusID:55096712} {\bibfield
  {journal} {\bibinfo  {journal} {Journal of High Energy Physics}\ }\textbf
  {\bibinfo {volume} {2018}} (\bibinfo {year} {2018})}\BibitemShut {NoStop}%
\bibitem [{\citenamefont {Zhou}\ and\ \citenamefont
  {Chen}(2019)}]{zhou2019operator}%
  \BibitemOpen
  \bibfield  {author} {\bibinfo {author} {\bibfnamefont {T.}~\bibnamefont
  {Zhou}}\ and\ \bibinfo {author} {\bibfnamefont {X.}~\bibnamefont {Chen}},\
  }\bibfield  {title} {\bibinfo {title} {{Operator dynamics in a Brownian
  quantum circuit}},\ }\href {https://doi.org/10.1103/PhysRevE.99.052212}
  {\bibfield  {journal} {\bibinfo  {journal} {Phys. Rev. E}\ }\textbf {\bibinfo
  {volume} {99}},\ \bibinfo {pages} {052212} (\bibinfo {year} {2019})},\
  \Eprint {https://arxiv.org/abs/1805.09307} {arXiv:1805.09307
  [cond-mat.str-el]} \BibitemShut {NoStop}%
\bibitem [{\citenamefont {Parker}\ \emph {et~al.}(2019)\citenamefont {Parker},
  \citenamefont {Cao}, \citenamefont {Avdoshkin}, \citenamefont {Scaffidi},\
  and\ \citenamefont {Altman}}]{parker2019A}%
  \BibitemOpen
  \bibfield  {author} {\bibinfo {author} {\bibfnamefont {D.~E.}\ \bibnamefont
  {Parker}}, \bibinfo {author} {\bibfnamefont {X.}~\bibnamefont {Cao}},
  \bibinfo {author} {\bibfnamefont {A.}~\bibnamefont {Avdoshkin}}, \bibinfo
  {author} {\bibfnamefont {T.}~\bibnamefont {Scaffidi}},\ and\ \bibinfo
  {author} {\bibfnamefont {E.}~\bibnamefont {Altman}},\ }\bibfield  {title}
  {\bibinfo {title} {A universal operator growth hypothesis},\ }\href
  {https://doi.org/10.1103/PhysRevX.9.041017} {\bibfield  {journal} {\bibinfo
  {journal} {Phys. Rev. X}\ }\textbf {\bibinfo {volume} {9}},\ \bibinfo {pages}
  {041017} (\bibinfo {year} {2019})}\BibitemShut {NoStop}%
\bibitem [{\citenamefont {Barb\'on}\ \emph {et~al.}(2019)\citenamefont
  {Barb\'on}, \citenamefont {Rabinovici}, \citenamefont {Shir},\ and\
  \citenamefont {Sinha}}]{barbon2019on}%
  \BibitemOpen
  \bibfield  {author} {\bibinfo {author} {\bibfnamefont {J.~L.~F.}\
  \bibnamefont {Barb\'on}}, \bibinfo {author} {\bibfnamefont {E.}~\bibnamefont
  {Rabinovici}}, \bibinfo {author} {\bibfnamefont {R.}~\bibnamefont {Shir}},\
  and\ \bibinfo {author} {\bibfnamefont {R.}~\bibnamefont {Sinha}},\ }\bibfield
   {title} {\bibinfo {title} {{On The Evolution Of Operator Complexity Beyond
  Scrambling}},\ }\href {https://doi.org/10.1007/JHEP10(2019)264} {\bibfield
  {journal} {\bibinfo  {journal} {JHEP}\ }\textbf {\bibinfo {volume} {10}},\
  \bibinfo {pages} {264}},\ \Eprint {https://arxiv.org/abs/1907.05393}
  {arXiv:1907.05393 [hep-th]} \BibitemShut {NoStop}%
\bibitem [{\citenamefont {Rabinovici}\ \emph {et~al.}(2021)\citenamefont
  {Rabinovici}, \citenamefont {S\'anchez-Garrido}, \citenamefont {Shir},\ and\
  \citenamefont {Sonner}}]{rabinovici2020operator}%
  \BibitemOpen
  \bibfield  {author} {\bibinfo {author} {\bibfnamefont {E.}~\bibnamefont
  {Rabinovici}}, \bibinfo {author} {\bibfnamefont {A.}~\bibnamefont
  {S\'anchez-Garrido}}, \bibinfo {author} {\bibfnamefont {R.}~\bibnamefont
  {Shir}},\ and\ \bibinfo {author} {\bibfnamefont {J.}~\bibnamefont {Sonner}},\
  }\bibfield  {title} {\bibinfo {title} {{Operator complexity: a journey to the
  edge of Krylov space}},\ }\href {https://doi.org/10.1007/JHEP06(2021)062}
  {\bibfield  {journal} {\bibinfo  {journal} {JHEP}\ }\textbf {\bibinfo
  {volume} {06}},\ \bibinfo {pages} {062}},\ \Eprint
  {https://arxiv.org/abs/2009.01862} {arXiv:2009.01862 [hep-th]} \BibitemShut
  {NoStop}%
\bibitem [{\citenamefont {Rabinovici}\ \emph
  {et~al.}(2022{\natexlab{a}})\citenamefont {Rabinovici}, \citenamefont
  {S\'anchez-Garrido}, \citenamefont {Shir},\ and\ \citenamefont
  {Sonner}}]{rabinovici2022krylov}%
  \BibitemOpen
  \bibfield  {author} {\bibinfo {author} {\bibfnamefont {E.}~\bibnamefont
  {Rabinovici}}, \bibinfo {author} {\bibfnamefont {A.}~\bibnamefont
  {S\'anchez-Garrido}}, \bibinfo {author} {\bibfnamefont {R.}~\bibnamefont
  {Shir}},\ and\ \bibinfo {author} {\bibfnamefont {J.}~\bibnamefont {Sonner}},\
  }\bibfield  {title} {\bibinfo {title} {{Krylov complexity from integrability
  to chaos}},\ }\href {https://doi.org/10.1007/JHEP07(2022)151} {\bibfield
  {journal} {\bibinfo  {journal} {JHEP}\ }\textbf {\bibinfo {volume} {07}},\
  \bibinfo {pages} {151}},\ \Eprint {https://arxiv.org/abs/2207.07701}
  {arXiv:2207.07701 [hep-th]} \BibitemShut {NoStop}%
\bibitem [{\citenamefont {Rabinovici}\ \emph
  {et~al.}(2022{\natexlab{b}})\citenamefont {Rabinovici}, \citenamefont
  {S\'anchez-Garrido}, \citenamefont {Shir},\ and\ \citenamefont
  {Sonner}}]{rabinovici2022krylov2}%
  \BibitemOpen
  \bibfield  {author} {\bibinfo {author} {\bibfnamefont {E.}~\bibnamefont
  {Rabinovici}}, \bibinfo {author} {\bibfnamefont {A.}~\bibnamefont
  {S\'anchez-Garrido}}, \bibinfo {author} {\bibfnamefont {R.}~\bibnamefont
  {Shir}},\ and\ \bibinfo {author} {\bibfnamefont {J.}~\bibnamefont {Sonner}},\
  }\bibfield  {title} {\bibinfo {title} {{Krylov localization and suppression
  of complexity}},\ }\href {https://doi.org/10.1007/JHEP03(2022)211} {\bibfield
   {journal} {\bibinfo  {journal} {JHEP}\ }\textbf {\bibinfo {volume} {03}},\
  \bibinfo {pages} {211}},\ \Eprint {https://arxiv.org/abs/2112.12128}
  {arXiv:2112.12128 [hep-th]} \BibitemShut {NoStop}%
\bibitem [{\citenamefont {Bhattacharjee}\ \emph
  {et~al.}(2022{\natexlab{a}})\citenamefont {Bhattacharjee}, \citenamefont
  {Cao}, \citenamefont {Nandy},\ and\ \citenamefont
  {Pathak}}]{bhattacharjee2022krylov}%
  \BibitemOpen
  \bibfield  {author} {\bibinfo {author} {\bibfnamefont {B.}~\bibnamefont
  {Bhattacharjee}}, \bibinfo {author} {\bibfnamefont {X.}~\bibnamefont {Cao}},
  \bibinfo {author} {\bibfnamefont {P.}~\bibnamefont {Nandy}},\ and\ \bibinfo
  {author} {\bibfnamefont {T.}~\bibnamefont {Pathak}},\ }\bibfield  {title}
  {\bibinfo {title} {{Krylov complexity in saddle-dominated scrambling}},\
  }\href {https://doi.org/10.1007/JHEP05(2022)174} {\bibfield  {journal}
  {\bibinfo  {journal} {JHEP}\ }\textbf {\bibinfo {volume} {05}},\ \bibinfo
  {pages} {174}},\ \Eprint {https://arxiv.org/abs/2203.03534} {arXiv:2203.03534
  [quant-ph]} \BibitemShut {NoStop}%
\bibitem [{\citenamefont {Bhattacharjee}\ \emph
  {et~al.}(2022{\natexlab{b}})\citenamefont {Bhattacharjee}, \citenamefont
  {Sur},\ and\ \citenamefont {Nandy}}]{bhattacharjee2022probing}%
  \BibitemOpen
  \bibfield  {author} {\bibinfo {author} {\bibfnamefont {B.}~\bibnamefont
  {Bhattacharjee}}, \bibinfo {author} {\bibfnamefont {S.}~\bibnamefont {Sur}},\
  and\ \bibinfo {author} {\bibfnamefont {P.}~\bibnamefont {Nandy}},\ }\bibfield
   {title} {\bibinfo {title} {{Probing quantum scars and weak ergodicity
  breaking through quantum complexity}},\ }\href
  {https://doi.org/10.1103/PhysRevB.106.205150} {\bibfield  {journal} {\bibinfo
   {journal} {Phys. Rev. B}\ }\textbf {\bibinfo {volume} {106}},\ \bibinfo
  {pages} {205150} (\bibinfo {year} {2022}{\natexlab{b}})},\ \Eprint
  {https://arxiv.org/abs/2208.05503} {arXiv:2208.05503 [quant-ph]} \BibitemShut
  {NoStop}%
\bibitem [{\citenamefont {Bhattacharjee}(2023)}]{bhattacharjee2023lanczos}%
  \BibitemOpen
  \bibfield  {author} {\bibinfo {author} {\bibfnamefont {B.}~\bibnamefont
  {Bhattacharjee}},\ }\href@noop {} {\bibinfo {title} {A lanczos approach to
  the adiabatic gauge potential}} (\bibinfo {year} {2023}),\ \Eprint
  {https://arxiv.org/abs/2302.07228} {arXiv:2302.07228 [quant-ph]} \BibitemShut
  {NoStop}%
\bibitem [{\citenamefont {Malishava}(2022)}]{malishava2022thermalization}%
  \BibitemOpen
  \bibfield  {author} {\bibinfo {author} {\bibfnamefont {M.}~\bibnamefont
  {Malishava}},\ }\href@noop {} {\bibinfo {title} {Thermalization of classical
  weakly nonintegrable many-body systems}} (\bibinfo {year} {2022}),\ \Eprint
  {https://arxiv.org/abs/2211.01887} {arXiv:2211.01887 [nlin.CD]} \BibitemShut
  {NoStop}%
\bibitem [{\citenamefont {Malishava}\ and\ \citenamefont
  {Flach}(2022)}]{malishava2022lyapunov}%
  \BibitemOpen
  \bibfield  {author} {\bibinfo {author} {\bibfnamefont {M.}~\bibnamefont
  {Malishava}}\ and\ \bibinfo {author} {\bibfnamefont {S.}~\bibnamefont
  {Flach}},\ }\bibfield  {title} {\bibinfo {title} {Lyapunov spectrum scaling
  for classical many-body dynamics close to integrability},\ }\href
  {https://doi.org/10.1103/PhysRevLett.128.134102} {\bibfield  {journal}
  {\bibinfo  {journal} {Phys. Rev. Lett.}\ }\textbf {\bibinfo {volume} {128}},\
  \bibinfo {pages} {134102} (\bibinfo {year} {2022})}\BibitemShut {NoStop}%
\bibitem [{\citenamefont {Lando}\ and\ \citenamefont
  {Flach}(2023)}]{lando2023thermalization}%
  \BibitemOpen
  \bibfield  {author} {\bibinfo {author} {\bibfnamefont {G.~M.}\ \bibnamefont
  {Lando}}\ and\ \bibinfo {author} {\bibfnamefont {S.}~\bibnamefont {Flach}},\
  }\bibfield  {title} {\bibinfo {title} {Thermalization slowing down in
  multidimensional josephson junction networks},\ }\href
  {https://doi.org/10.1103/PhysRevE.108.L062301} {\bibfield  {journal}
  {\bibinfo  {journal} {Phys. Rev. E}\ }\textbf {\bibinfo {volume} {108}},\
  \bibinfo {pages} {L062301} (\bibinfo {year} {2023})}\BibitemShut {NoStop}%
\bibitem [{\citenamefont {Zhang}\ \emph {et~al.}(2024)\citenamefont {Zhang},
  \citenamefont {Lando}, \citenamefont {Dietz},\ and\ \citenamefont
  {Flach}}]{zhang2024thermalization}%
  \BibitemOpen
  \bibfield  {author} {\bibinfo {author} {\bibfnamefont {W.}~\bibnamefont
  {Zhang}}, \bibinfo {author} {\bibfnamefont {G.~M.}\ \bibnamefont {Lando}},
  \bibinfo {author} {\bibfnamefont {B.}~\bibnamefont {Dietz}},\ and\ \bibinfo
  {author} {\bibfnamefont {S.}~\bibnamefont {Flach}},\ }\bibfield  {title}
  {\bibinfo {title} {Thermalization universality-class transition induced by
  anderson localization},\ }\href
  {https://doi.org/10.1103/PhysRevResearch.6.L012064} {\bibfield  {journal}
  {\bibinfo  {journal} {Phys. Rev. Res.}\ }\textbf {\bibinfo {volume} {6}},\
  \bibinfo {pages} {L012064} (\bibinfo {year} {2024})}\BibitemShut {NoStop}%
\bibitem [{\citenamefont {Danieli}\ \emph {et~al.}(2017)\citenamefont
  {Danieli}, \citenamefont {Campbell},\ and\ \citenamefont
  {Flach}}]{danieli2017intermittent}%
  \BibitemOpen
  \bibfield  {author} {\bibinfo {author} {\bibfnamefont {C.}~\bibnamefont
  {Danieli}}, \bibinfo {author} {\bibfnamefont {D.}~\bibnamefont {Campbell}},\
  and\ \bibinfo {author} {\bibfnamefont {S.}~\bibnamefont {Flach}},\ }\bibfield
   {title} {\bibinfo {title} {Intermittent many-body dynamics at equilibrium},\
  }\href {https://doi.org/10.1103/PhysRevE.95.060202} {\bibfield  {journal}
  {\bibinfo  {journal} {Phys. Rev. E}\ }\textbf {\bibinfo {volume} {95}},\
  \bibinfo {pages} {060202} (\bibinfo {year} {2017})}\BibitemShut {NoStop}%
\bibitem [{\citenamefont {Danieli}\ \emph {et~al.}(2019)\citenamefont
  {Danieli}, \citenamefont {Mithun}, \citenamefont {Kati}, \citenamefont
  {Campbell},\ and\ \citenamefont {Flach}}]{danieli2019dynamical}%
  \BibitemOpen
  \bibfield  {author} {\bibinfo {author} {\bibfnamefont {C.}~\bibnamefont
  {Danieli}}, \bibinfo {author} {\bibfnamefont {T.}~\bibnamefont {Mithun}},
  \bibinfo {author} {\bibfnamefont {Y.}~\bibnamefont {Kati}}, \bibinfo {author}
  {\bibfnamefont {D.~K.}\ \bibnamefont {Campbell}},\ and\ \bibinfo {author}
  {\bibfnamefont {S.}~\bibnamefont {Flach}},\ }\bibfield  {title} {\bibinfo
  {title} {Dynamical glass in weakly nonintegrable klein-gordon chains},\
  }\href {https://doi.org/10.1103/PhysRevE.100.032217} {\bibfield  {journal}
  {\bibinfo  {journal} {Phys. Rev. E}\ }\textbf {\bibinfo {volume} {100}},\
  \bibinfo {pages} {032217} (\bibinfo {year} {2019})}\BibitemShut {NoStop}%
\bibitem [{\citenamefont {Mithun}\ \emph {et~al.}(2018)\citenamefont {Mithun},
  \citenamefont {Kati}, \citenamefont {Danieli},\ and\ \citenamefont
  {Flach}}]{mithun2018weakly}%
  \BibitemOpen
  \bibfield  {author} {\bibinfo {author} {\bibfnamefont {T.}~\bibnamefont
  {Mithun}}, \bibinfo {author} {\bibfnamefont {Y.}~\bibnamefont {Kati}},
  \bibinfo {author} {\bibfnamefont {C.}~\bibnamefont {Danieli}},\ and\ \bibinfo
  {author} {\bibfnamefont {S.}~\bibnamefont {Flach}},\ }\bibfield  {title}
  {\bibinfo {title} {Weakly nonergodic dynamics in the gross-pitaevskii
  lattice},\ }\href {https://doi.org/10.1103/PhysRevLett.120.184101} {\bibfield
   {journal} {\bibinfo  {journal} {Phys. Rev. Lett.}\ }\textbf {\bibinfo
  {volume} {120}},\ \bibinfo {pages} {184101} (\bibinfo {year}
  {2018})}\BibitemShut {NoStop}%
\bibitem [{\citenamefont {Mithun}\ \emph {et~al.}(2019)\citenamefont {Mithun},
  \citenamefont {Danieli}, \citenamefont {Kati},\ and\ \citenamefont
  {Flach}}]{mithun2019dynamical}%
  \BibitemOpen
  \bibfield  {author} {\bibinfo {author} {\bibfnamefont {T.}~\bibnamefont
  {Mithun}}, \bibinfo {author} {\bibfnamefont {C.}~\bibnamefont {Danieli}},
  \bibinfo {author} {\bibfnamefont {Y.}~\bibnamefont {Kati}},\ and\ \bibinfo
  {author} {\bibfnamefont {S.}~\bibnamefont {Flach}},\ }\bibfield  {title}
  {\bibinfo {title} {Dynamical glass and ergodization times in classical
  josephson junction chains},\ }\href
  {https://doi.org/10.1103/PhysRevLett.122.054102} {\bibfield  {journal}
  {\bibinfo  {journal} {Phys. Rev. Lett.}\ }\textbf {\bibinfo {volume} {122}},\
  \bibinfo {pages} {054102} (\bibinfo {year} {2019})}\BibitemShut {NoStop}%
\bibitem [{\citenamefont {Bel}\ and\ \citenamefont
  {Barkai}(2005)}]{bel2005weak}%
  \BibitemOpen
  \bibfield  {author} {\bibinfo {author} {\bibfnamefont {G.}~\bibnamefont
  {Bel}}\ and\ \bibinfo {author} {\bibfnamefont {E.}~\bibnamefont {Barkai}},\
  }\bibfield  {title} {\bibinfo {title} {Weak ergodicity breaking in the
  continuous-time random walk},\ }\href
  {https://doi.org/10.1103/PhysRevLett.94.240602} {\bibfield  {journal}
  {\bibinfo  {journal} {Phys. Rev. Lett.}\ }\textbf {\bibinfo {volume} {94}},\
  \bibinfo {pages} {240602} (\bibinfo {year} {2005})}\BibitemShut {NoStop}%
\bibitem [{\citenamefont {Rigol}(2009)}]{rigol2009breakdown}%
  \BibitemOpen
  \bibfield  {author} {\bibinfo {author} {\bibfnamefont {M.}~\bibnamefont
  {Rigol}},\ }\bibfield  {title} {\bibinfo {title} {Breakdown of thermalization
  in finite one-dimensional systems},\ }\href
  {https://doi.org/10.1103/PhysRevLett.103.100403} {\bibfield  {journal}
  {\bibinfo  {journal} {Phys. Rev. Lett.}\ }\textbf {\bibinfo {volume} {103}},\
  \bibinfo {pages} {100403} (\bibinfo {year} {2009})}\BibitemShut {NoStop}%
\bibitem [{\citenamefont {Srednicki}(1994)}]{srednicki1994chaos}%
  \BibitemOpen
  \bibfield  {author} {\bibinfo {author} {\bibfnamefont {M.}~\bibnamefont
  {Srednicki}},\ }\bibfield  {title} {\bibinfo {title} {Chaos and quantum
  thermalization},\ }\href {https://doi.org/10.1103/PhysRevE.50.888} {\bibfield
   {journal} {\bibinfo  {journal} {Phys. Rev. E}\ }\textbf {\bibinfo {volume}
  {50}},\ \bibinfo {pages} {888} (\bibinfo {year} {1994})}\BibitemShut
  {NoStop}%
\bibitem [{\citenamefont {Deutsch}(1991)}]{deutsch1991quantum}%
  \BibitemOpen
  \bibfield  {author} {\bibinfo {author} {\bibfnamefont {J.~M.}\ \bibnamefont
  {Deutsch}},\ }\bibfield  {title} {\bibinfo {title} {Quantum statistical
  mechanics in a closed system},\ }\href
  {https://doi.org/10.1103/PhysRevA.43.2046} {\bibfield  {journal} {\bibinfo
  {journal} {Phys. Rev. A}\ }\textbf {\bibinfo {volume} {43}},\ \bibinfo
  {pages} {2046} (\bibinfo {year} {1991})}\BibitemShut {NoStop}%
\bibitem [{\citenamefont {D'Alessio}\ \emph {et~al.}(2016)\citenamefont
  {D'Alessio}, \citenamefont {Kafri}, \citenamefont {Polkovnikov},\ and\
  \citenamefont {Rigol}}]{dalessio2015from}%
  \BibitemOpen
  \bibfield  {author} {\bibinfo {author} {\bibfnamefont {L.}~\bibnamefont
  {D'Alessio}}, \bibinfo {author} {\bibfnamefont {Y.}~\bibnamefont {Kafri}},
  \bibinfo {author} {\bibfnamefont {A.}~\bibnamefont {Polkovnikov}},\ and\
  \bibinfo {author} {\bibfnamefont {M.}~\bibnamefont {Rigol}},\ }\bibfield
  {title} {\bibinfo {title} {{From quantum chaos and eigenstate thermalization
  to statistical mechanics and thermodynamics}},\ }\href
  {https://doi.org/10.1080/00018732.2016.1198134} {\bibfield  {journal}
  {\bibinfo  {journal} {Adv. Phys.}\ }\textbf {\bibinfo {volume} {65}},\
  \bibinfo {pages} {239} (\bibinfo {year} {2016})},\ \Eprint
  {https://arxiv.org/abs/1509.06411} {arXiv:1509.06411 [cond-mat.stat-mech]}
  \BibitemShut {NoStop}%
\bibitem [{\citenamefont {Magan}(2016)}]{magan2016random}%
  \BibitemOpen
  \bibfield  {author} {\bibinfo {author} {\bibfnamefont {J.~M.}\ \bibnamefont
  {Magan}},\ }\bibfield  {title} {\bibinfo {title} {{Random free fermions: An
  analytical example of eigenstate thermalization}},\ }\href
  {https://doi.org/10.1103/PhysRevLett.116.030401} {\bibfield  {journal}
  {\bibinfo  {journal} {Phys. Rev. Lett.}\ }\textbf {\bibinfo {volume} {116}},\
  \bibinfo {pages} {030401} (\bibinfo {year} {2016})},\ \Eprint
  {https://arxiv.org/abs/1508.05339} {arXiv:1508.05339 [quant-ph]} \BibitemShut
  {NoStop}%
\bibitem [{\citenamefont {Deutsch}(2018)}]{deutsch2018eigenstate}%
  \BibitemOpen
  \bibfield  {author} {\bibinfo {author} {\bibfnamefont {J.~M.}\ \bibnamefont
  {Deutsch}},\ }\bibfield  {title} {\bibinfo {title} {Eigenstate thermalization
  hypothesis},\ }\href {https://doi.org/10.1088/1361-6633/aac9f1} {\bibfield
  {journal} {\bibinfo  {journal} {Reports on Progress in Physics}\ }\textbf
  {\bibinfo {volume} {81}},\ \bibinfo {pages} {082001} (\bibinfo {year}
  {2018})}\BibitemShut {NoStop}%
\bibitem [{\citenamefont {Brenes}\ \emph
  {et~al.}(2020{\natexlab{a}})\citenamefont {Brenes}, \citenamefont {LeBlond},
  \citenamefont {Goold},\ and\ \citenamefont {Rigol}}]{brenes2020eigenstate}%
  \BibitemOpen
  \bibfield  {author} {\bibinfo {author} {\bibfnamefont {M.}~\bibnamefont
  {Brenes}}, \bibinfo {author} {\bibfnamefont {T.}~\bibnamefont {LeBlond}},
  \bibinfo {author} {\bibfnamefont {J.}~\bibnamefont {Goold}},\ and\ \bibinfo
  {author} {\bibfnamefont {M.}~\bibnamefont {Rigol}},\ }\bibfield  {title}
  {\bibinfo {title} {Eigenstate thermalization in a locally perturbed
  integrable system},\ }\href {https://doi.org/10.1103/PhysRevLett.125.070605}
  {\bibfield  {journal} {\bibinfo  {journal} {Phys. Rev. Lett.}\ }\textbf
  {\bibinfo {volume} {125}},\ \bibinfo {pages} {070605} (\bibinfo {year}
  {2020}{\natexlab{a}})}\BibitemShut {NoStop}%
\bibitem [{\citenamefont {Jafferis}\ \emph {et~al.}(2023)\citenamefont
  {Jafferis}, \citenamefont {Kolchmeyer}, \citenamefont {Mukhametzhanov},\ and\
  \citenamefont {Sonner}}]{jafferis2023matrix}%
  \BibitemOpen
  \bibfield  {author} {\bibinfo {author} {\bibfnamefont {D.~L.}\ \bibnamefont
  {Jafferis}}, \bibinfo {author} {\bibfnamefont {D.~K.}\ \bibnamefont
  {Kolchmeyer}}, \bibinfo {author} {\bibfnamefont {B.}~\bibnamefont
  {Mukhametzhanov}},\ and\ \bibinfo {author} {\bibfnamefont {J.}~\bibnamefont
  {Sonner}},\ }\href@noop {} {\bibinfo {title} {Matrix models for eigenstate
  thermalization}} (\bibinfo {year} {2023}),\ \Eprint
  {https://arxiv.org/abs/2209.02130} {arXiv:2209.02130 [hep-th]} \BibitemShut
  {NoStop}%
\bibitem [{\citenamefont {Wang}\ \emph {et~al.}(2022)\citenamefont {Wang},
  \citenamefont {Lamann}, \citenamefont {Richter}, \citenamefont {Steinigeweg},
  \citenamefont {Dymarsky},\ and\ \citenamefont {Gemmer}}]{Wang2021eigenstate}%
  \BibitemOpen
  \bibfield  {author} {\bibinfo {author} {\bibfnamefont {J.}~\bibnamefont
  {Wang}}, \bibinfo {author} {\bibfnamefont {M.~H.}\ \bibnamefont {Lamann}},
  \bibinfo {author} {\bibfnamefont {J.}~\bibnamefont {Richter}}, \bibinfo
  {author} {\bibfnamefont {R.}~\bibnamefont {Steinigeweg}}, \bibinfo {author}
  {\bibfnamefont {A.}~\bibnamefont {Dymarsky}},\ and\ \bibinfo {author}
  {\bibfnamefont {J.}~\bibnamefont {Gemmer}},\ }\bibfield  {title} {\bibinfo
  {title} {{Eigenstate Thermalization Hypothesis and Its Deviations from
  Random-Matrix Theory beyond the Thermalization Time}},\ }\href
  {https://doi.org/10.1103/PhysRevLett.128.180601} {\bibfield  {journal}
  {\bibinfo  {journal} {Phys. Rev. Lett.}\ }\textbf {\bibinfo {volume} {128}},\
  \bibinfo {pages} {180601} (\bibinfo {year} {2022})},\ \Eprint
  {https://arxiv.org/abs/2110.04085} {arXiv:2110.04085 [cond-mat.stat-mech]}
  \BibitemShut {NoStop}%
\bibitem [{\citenamefont {Dymarsky}(2022)}]{dymarsky2022bound}%
  \BibitemOpen
  \bibfield  {author} {\bibinfo {author} {\bibfnamefont {A.}~\bibnamefont
  {Dymarsky}},\ }\bibfield  {title} {\bibinfo {title} {{Bound on Eigenstate
  Thermalization from Transport}},\ }\href
  {https://doi.org/10.1103/PhysRevLett.128.190601} {\bibfield  {journal}
  {\bibinfo  {journal} {Phys. Rev. Lett.}\ }\textbf {\bibinfo {volume} {128}},\
  \bibinfo {pages} {190601} (\bibinfo {year} {2022})},\ \Eprint
  {https://arxiv.org/abs/1804.08626} {arXiv:1804.08626 [cond-mat.stat-mech]}
  \BibitemShut {NoStop}%
\bibitem [{\citenamefont {Lezama}\ \emph {et~al.}(2021)\citenamefont {Lezama},
  \citenamefont {Torres-Herrera}, \citenamefont {P\'erez-Bernal}, \citenamefont
  {Bar~Lev},\ and\ \citenamefont {Santos}}]{lezama2021equilibration}%
  \BibitemOpen
  \bibfield  {author} {\bibinfo {author} {\bibfnamefont {T.~L.~M.}\
  \bibnamefont {Lezama}}, \bibinfo {author} {\bibfnamefont {E.~J.}\
  \bibnamefont {Torres-Herrera}}, \bibinfo {author} {\bibfnamefont
  {F.}~\bibnamefont {P\'erez-Bernal}}, \bibinfo {author} {\bibfnamefont
  {Y.}~\bibnamefont {Bar~Lev}},\ and\ \bibinfo {author} {\bibfnamefont {L.~F.}\
  \bibnamefont {Santos}},\ }\bibfield  {title} {\bibinfo {title} {Equilibration
  time in many-body quantum systems},\ }\href
  {https://doi.org/10.1103/PhysRevB.104.085117} {\bibfield  {journal} {\bibinfo
   {journal} {Phys. Rev. B}\ }\textbf {\bibinfo {volume} {104}},\ \bibinfo
  {pages} {085117} (\bibinfo {year} {2021})}\BibitemShut {NoStop}%
\bibitem [{\citenamefont {Schiulaz}\ \emph {et~al.}(2015)\citenamefont
  {Schiulaz}, \citenamefont {Silva},\ and\ \citenamefont
  {M\"uller}}]{schiulaz2015dynamics}%
  \BibitemOpen
  \bibfield  {author} {\bibinfo {author} {\bibfnamefont {M.}~\bibnamefont
  {Schiulaz}}, \bibinfo {author} {\bibfnamefont {A.}~\bibnamefont {Silva}},\
  and\ \bibinfo {author} {\bibfnamefont {M.}~\bibnamefont {M\"uller}},\
  }\bibfield  {title} {\bibinfo {title} {Dynamics in many-body localized
  quantum systems without disorder},\ }\href
  {https://doi.org/10.1103/PhysRevB.91.184202} {\bibfield  {journal} {\bibinfo
  {journal} {Phys. Rev. B}\ }\textbf {\bibinfo {volume} {91}},\ \bibinfo
  {pages} {184202} (\bibinfo {year} {2015})}\BibitemShut {NoStop}%
\bibitem [{\citenamefont {Rozenbaum}\ \emph {et~al.}(2017)\citenamefont
  {Rozenbaum}, \citenamefont {Ganeshan},\ and\ \citenamefont
  {Galitski}}]{rozenbaum2016lyapunov}%
  \BibitemOpen
  \bibfield  {author} {\bibinfo {author} {\bibfnamefont {E.~B.}\ \bibnamefont
  {Rozenbaum}}, \bibinfo {author} {\bibfnamefont {S.}~\bibnamefont
  {Ganeshan}},\ and\ \bibinfo {author} {\bibfnamefont {V.}~\bibnamefont
  {Galitski}},\ }\bibfield  {title} {\bibinfo {title} {{Lyapunov Exponent and
  Out-of-Time-Ordered Correlator\textquoteright{}s Growth Rate in a Chaotic
  System}},\ }\href {https://doi.org/10.1103/PhysRevLett.118.086801} {\bibfield
   {journal} {\bibinfo  {journal} {Phys. Rev. Lett.}\ }\textbf {\bibinfo
  {volume} {118}},\ \bibinfo {pages} {086801} (\bibinfo {year} {2017})},\
  \Eprint {https://arxiv.org/abs/1609.01707} {arXiv:1609.01707
  [cond-mat.dis-nn]} \BibitemShut {NoStop}%
\bibitem [{\citenamefont {Hallam}\ \emph {et~al.}(2019)\citenamefont {Hallam},
  \citenamefont {Morley},\ and\ \citenamefont {Green}}]{hallam2018the}%
  \BibitemOpen
  \bibfield  {author} {\bibinfo {author} {\bibfnamefont {A.}~\bibnamefont
  {Hallam}}, \bibinfo {author} {\bibfnamefont {J.~G.}\ \bibnamefont {Morley}},\
  and\ \bibinfo {author} {\bibfnamefont {A.~G.}\ \bibnamefont {Green}},\
  }\bibfield  {title} {\bibinfo {title} {The lyapunov spectra of quantum
  thermalisation},\ }\href {https://doi.org/10.1038/s41467-019-10336-4}
  {\bibfield  {journal} {\bibinfo  {journal} {Nature Communications}\ }\textbf
  {\bibinfo {volume} {10}},\ \bibinfo {pages} {2708} (\bibinfo {year}
  {2019})},\ \Eprint {https://arxiv.org/abs/1806.05204} {arXiv:1806.05204
  [cond-mat.str-el]} \BibitemShut {NoStop}%
\bibitem [{\citenamefont {Gharibyan}\ \emph {et~al.}(2019)\citenamefont
  {Gharibyan}, \citenamefont {Hanada}, \citenamefont {Swingle},\ and\
  \citenamefont {Tezuka}}]{gharibyan2018quantum}%
  \BibitemOpen
  \bibfield  {author} {\bibinfo {author} {\bibfnamefont {H.}~\bibnamefont
  {Gharibyan}}, \bibinfo {author} {\bibfnamefont {M.}~\bibnamefont {Hanada}},
  \bibinfo {author} {\bibfnamefont {B.}~\bibnamefont {Swingle}},\ and\ \bibinfo
  {author} {\bibfnamefont {M.}~\bibnamefont {Tezuka}},\ }\bibfield  {title}
  {\bibinfo {title} {Quantum lyapunov spectrum},\ }\href
  {https://doi.org/10.1007/JHEP04(2019)082} {\bibfield  {journal} {\bibinfo
  {journal} {Journal of High Energy Physics}\ }\textbf {\bibinfo {volume}
  {2019}},\ \bibinfo {pages} {82} (\bibinfo {year} {2019})},\ \Eprint
  {https://arxiv.org/abs/1809.01671} {arXiv:1809.01671 [quant-ph]} \BibitemShut
  {NoStop}%
\bibitem [{\citenamefont {Chan}\ \emph {et~al.}(2021)\citenamefont {Chan},
  \citenamefont {De~Luca},\ and\ \citenamefont {Chalker}}]{chan2021spectral}%
  \BibitemOpen
  \bibfield  {author} {\bibinfo {author} {\bibfnamefont {A.}~\bibnamefont
  {Chan}}, \bibinfo {author} {\bibfnamefont {A.}~\bibnamefont {De~Luca}},\ and\
  \bibinfo {author} {\bibfnamefont {J.~T.}\ \bibnamefont {Chalker}},\
  }\bibfield  {title} {\bibinfo {title} {Spectral lyapunov exponents in chaotic
  and localized many-body quantum systems},\ }\href
  {https://doi.org/10.1103/PhysRevResearch.3.023118} {\bibfield  {journal}
  {\bibinfo  {journal} {Phys. Rev. Res.}\ }\textbf {\bibinfo {volume} {3}},\
  \bibinfo {pages} {023118} (\bibinfo {year} {2021})}\BibitemShut {NoStop}%
\bibitem [{\citenamefont {Maldacena}\ \emph {et~al.}(2016)\citenamefont
  {Maldacena}, \citenamefont {Shenker},\ and\ \citenamefont
  {Stanford}}]{maldacena2015A}%
  \BibitemOpen
  \bibfield  {author} {\bibinfo {author} {\bibfnamefont {J.}~\bibnamefont
  {Maldacena}}, \bibinfo {author} {\bibfnamefont {S.~H.}\ \bibnamefont
  {Shenker}},\ and\ \bibinfo {author} {\bibfnamefont {D.}~\bibnamefont
  {Stanford}},\ }\bibfield  {title} {\bibinfo {title} {{A bound on chaos}},\
  }\href {https://doi.org/10.1007/JHEP08(2016)106} {\bibfield  {journal}
  {\bibinfo  {journal} {JHEP}\ }\textbf {\bibinfo {volume} {08}},\ \bibinfo
  {pages} {106}},\ \Eprint {https://arxiv.org/abs/1503.01409} {arXiv:1503.01409
  [hep-th]} \BibitemShut {NoStop}%
\bibitem [{\citenamefont {Pfeuty}(1970)}]{pfeuty1970the}%
  \BibitemOpen
  \bibfield  {author} {\bibinfo {author} {\bibfnamefont {P.}~\bibnamefont
  {Pfeuty}},\ }\bibfield  {title} {\bibinfo {title} {The one-dimensional ising
  model with a transverse field},\ }\href
  {https://doi.org/https://doi.org/10.1016/0003-4916(70)90270-8} {\bibfield
  {journal} {\bibinfo  {journal} {Annals of Physics}\ }\textbf {\bibinfo
  {volume} {57}},\ \bibinfo {pages} {79} (\bibinfo {year} {1970})}\BibitemShut
  {NoStop}%
\bibitem [{\citenamefont {Sachdev}(2007)}]{sachdev2007quantum}%
  \BibitemOpen
  \bibfield  {author} {\bibinfo {author} {\bibfnamefont {S.}~\bibnamefont
  {Sachdev}},\ }\href {https://doi.org/10.1002/9780470022184.hmm108} {\emph
  {\bibinfo {title} {Handbook of Magnetism and Advanced Magnetic Materials}}}\
  (\bibinfo  {publisher} {John Wiley and Sons, Ltd},\ \bibinfo {year}
  {2007})\BibitemShut {NoStop}%
\bibitem [{\citenamefont {Ba\~nuls}\ \emph {et~al.}(2011)\citenamefont
  {Ba\~nuls}, \citenamefont {Cirac},\ and\ \citenamefont
  {Hastings}}]{banuls2011strong}%
  \BibitemOpen
  \bibfield  {author} {\bibinfo {author} {\bibfnamefont {M.~C.}\ \bibnamefont
  {Ba\~nuls}}, \bibinfo {author} {\bibfnamefont {J.~I.}\ \bibnamefont
  {Cirac}},\ and\ \bibinfo {author} {\bibfnamefont {M.~B.}\ \bibnamefont
  {Hastings}},\ }\bibfield  {title} {\bibinfo {title} {Strong and weak
  thermalization of infinite nonintegrable quantum systems},\ }\href
  {https://doi.org/10.1103/PhysRevLett.106.050405} {\bibfield  {journal}
  {\bibinfo  {journal} {Phys. Rev. Lett.}\ }\textbf {\bibinfo {volume} {106}},\
  \bibinfo {pages} {050405} (\bibinfo {year} {2011})}\BibitemShut {NoStop}%
\bibitem [{\citenamefont {Roberts}\ \emph {et~al.}(2015)\citenamefont
  {Roberts}, \citenamefont {Stanford},\ and\ \citenamefont
  {Susskind}}]{roberts2014localized}%
  \BibitemOpen
  \bibfield  {author} {\bibinfo {author} {\bibfnamefont {D.~A.}\ \bibnamefont
  {Roberts}}, \bibinfo {author} {\bibfnamefont {D.}~\bibnamefont {Stanford}},\
  and\ \bibinfo {author} {\bibfnamefont {L.}~\bibnamefont {Susskind}},\
  }\bibfield  {title} {\bibinfo {title} {{Localized shocks}},\ }\href
  {https://doi.org/10.1007/JHEP03(2015)051} {\bibfield  {journal} {\bibinfo
  {journal} {JHEP}\ }\textbf {\bibinfo {volume} {03}},\ \bibinfo {pages}
  {051}},\ \Eprint {https://arxiv.org/abs/1409.8180} {arXiv:1409.8180 [hep-th]}
  \BibitemShut {NoStop}%
\bibitem [{\citenamefont {Lakshminarayan}\ and\ \citenamefont
  {Subrahmanyam}(2005)}]{arul2005multipartite}%
  \BibitemOpen
  \bibfield  {author} {\bibinfo {author} {\bibfnamefont {A.}~\bibnamefont
  {Lakshminarayan}}\ and\ \bibinfo {author} {\bibfnamefont {V.}~\bibnamefont
  {Subrahmanyam}},\ }\bibfield  {title} {\bibinfo {title} {Multipartite
  entanglement in a one-dimensional time-dependent ising model},\ }\href
  {https://doi.org/10.1103/PhysRevA.71.062334} {\bibfield  {journal} {\bibinfo
  {journal} {Phys. Rev. A}\ }\textbf {\bibinfo {volume} {71}},\ \bibinfo
  {pages} {062334} (\bibinfo {year} {2005})}\BibitemShut {NoStop}%
\bibitem [{\citenamefont {Lin}\ and\ \citenamefont
  {Motrunich}(2018)}]{lin2018otoc}%
  \BibitemOpen
  \bibfield  {author} {\bibinfo {author} {\bibfnamefont {C.-J.}\ \bibnamefont
  {Lin}}\ and\ \bibinfo {author} {\bibfnamefont {O.~I.}\ \bibnamefont
  {Motrunich}},\ }\bibfield  {title} {\bibinfo {title} {Out-of-time-ordered
  correlators in a quantum ising chain},\ }\href
  {https://doi.org/10.1103/PhysRevB.97.144304} {\bibfield  {journal} {\bibinfo
  {journal} {Phys. Rev. B}\ }\textbf {\bibinfo {volume} {97}},\ \bibinfo
  {pages} {144304} (\bibinfo {year} {2018})}\BibitemShut {NoStop}%
\bibitem [{\citenamefont {Craps}\ \emph {et~al.}(2020)\citenamefont {Craps},
  \citenamefont {De~Clerck}, \citenamefont {Janssens}, \citenamefont {Luyten},\
  and\ \citenamefont {Rabideau}}]{craps2020lyapunov}%
  \BibitemOpen
  \bibfield  {author} {\bibinfo {author} {\bibfnamefont {B.}~\bibnamefont
  {Craps}}, \bibinfo {author} {\bibfnamefont {M.}~\bibnamefont {De~Clerck}},
  \bibinfo {author} {\bibfnamefont {D.}~\bibnamefont {Janssens}}, \bibinfo
  {author} {\bibfnamefont {V.}~\bibnamefont {Luyten}},\ and\ \bibinfo {author}
  {\bibfnamefont {C.}~\bibnamefont {Rabideau}},\ }\bibfield  {title} {\bibinfo
  {title} {Lyapunov growth in quantum spin chains},\ }\href
  {https://doi.org/10.1103/PhysRevB.101.174313} {\bibfield  {journal} {\bibinfo
   {journal} {Phys. Rev. B}\ }\textbf {\bibinfo {volume} {101}},\ \bibinfo
  {pages} {174313} (\bibinfo {year} {2020})}\BibitemShut {NoStop}%
\bibitem [{\citenamefont {Nivedita}\ \emph {et~al.}(2020)\citenamefont
  {Nivedita}, \citenamefont {Shackleton},\ and\ \citenamefont
  {Sachdev}}]{nivedita2020spectral}%
  \BibitemOpen
  \bibfield  {author} {\bibinfo {author} {\bibnamefont {Nivedita}}, \bibinfo
  {author} {\bibfnamefont {H.}~\bibnamefont {Shackleton}},\ and\ \bibinfo
  {author} {\bibfnamefont {S.}~\bibnamefont {Sachdev}},\ }\bibfield  {title}
  {\bibinfo {title} {Spectral form factors of clean and random quantum ising
  chains},\ }\href {https://doi.org/10.1103/PhysRevE.101.042136} {\bibfield
  {journal} {\bibinfo  {journal} {Phys. Rev. E}\ }\textbf {\bibinfo {volume}
  {101}},\ \bibinfo {pages} {042136} (\bibinfo {year} {2020})}\BibitemShut
  {NoStop}%
\bibitem [{Note1()}]{Note1}%
  \BibitemOpen
  \bibinfo {note} {Corresponding to the fermionic operators in the
  Jordan-Wigner representation of the system}\BibitemShut {NoStop}%
\bibitem [{Note2()}]{Note2}%
  \BibitemOpen
  \bibinfo {note} {Somewhat related to this is the delocalization in Fock
  space, which has been studied in Ref.~\protect \rev@citealp
  {bulchandani2022onset}}\BibitemShut {NoStop}%
\bibitem [{Note3()}]{Note3}%
  \BibitemOpen
  \bibinfo {note} {Minimal with respect to a cost function; which in this case
  is the Krylov complexity itself.}\BibitemShut {Stop}%
\bibitem [{\citenamefont {Viswanath}\ and\ \citenamefont
  {M{\"u}ller}(1994)}]{viswanath1994recursion}%
  \BibitemOpen
  \bibfield  {author} {\bibinfo {author} {\bibfnamefont {V.}~\bibnamefont
  {Viswanath}}\ and\ \bibinfo {author} {\bibfnamefont {G.}~\bibnamefont
  {M{\"u}ller}},\ }\href {https://books.google.co.in/books?id=X2Ug4w17rnMC}
  {\emph {\bibinfo {title} {The Recursion Method: Application to Many Body
  Dynamics}}},\ Lecture Notes in Physics Monographs\ (\bibinfo  {publisher}
  {Springer Berlin Heidelberg},\ \bibinfo {year} {1994})\BibitemShut {NoStop}%
\bibitem [{\citenamefont {Zhang}\ \emph {et~al.}(2022)\citenamefont {Zhang},
  \citenamefont {Vidmar},\ and\ \citenamefont {Rigol}}]{zhang2022statistical}%
  \BibitemOpen
  \bibfield  {author} {\bibinfo {author} {\bibfnamefont {Y.}~\bibnamefont
  {Zhang}}, \bibinfo {author} {\bibfnamefont {L.}~\bibnamefont {Vidmar}},\ and\
  \bibinfo {author} {\bibfnamefont {M.}~\bibnamefont {Rigol}},\ }\bibfield
  {title} {\bibinfo {title} {Statistical properties of the off-diagonal matrix
  elements of observables in eigenstates of integrable systems},\ }\href
  {https://doi.org/10.1103/PhysRevE.106.014132} {\bibfield  {journal} {\bibinfo
   {journal} {Phys. Rev. E}\ }\textbf {\bibinfo {volume} {106}},\ \bibinfo
  {pages} {014132} (\bibinfo {year} {2022})}\BibitemShut {NoStop}%
\bibitem [{\citenamefont {Richter}\ \emph {et~al.}(2020)\citenamefont
  {Richter}, \citenamefont {Dymarsky}, \citenamefont {Steinigeweg},\ and\
  \citenamefont {Gemmer}}]{richter2020eigenstate}%
  \BibitemOpen
  \bibfield  {author} {\bibinfo {author} {\bibfnamefont {J.}~\bibnamefont
  {Richter}}, \bibinfo {author} {\bibfnamefont {A.}~\bibnamefont {Dymarsky}},
  \bibinfo {author} {\bibfnamefont {R.}~\bibnamefont {Steinigeweg}},\ and\
  \bibinfo {author} {\bibfnamefont {J.}~\bibnamefont {Gemmer}},\ }\bibfield
  {title} {\bibinfo {title} {Eigenstate thermalization hypothesis beyond
  standard indicators: Emergence of random-matrix behavior at small
  frequencies},\ }\href {https://doi.org/10.1103/PhysRevE.102.042127}
  {\bibfield  {journal} {\bibinfo  {journal} {Phys. Rev. E}\ }\textbf {\bibinfo
  {volume} {102}},\ \bibinfo {pages} {042127} (\bibinfo {year}
  {2020})}\BibitemShut {NoStop}%
\bibitem [{\citenamefont {Fritzsch}\ and\ \citenamefont
  {Prosen}(2021)}]{frizsch2021eigenstate}%
  \BibitemOpen
  \bibfield  {author} {\bibinfo {author} {\bibfnamefont {F.}~\bibnamefont
  {Fritzsch}}\ and\ \bibinfo {author} {\bibfnamefont {T.~c.~v.}\ \bibnamefont
  {Prosen}},\ }\bibfield  {title} {\bibinfo {title} {Eigenstate thermalization
  in dual-unitary quantum circuits: Asymptotics of spectral functions},\ }\href
  {https://doi.org/10.1103/PhysRevE.103.062133} {\bibfield  {journal} {\bibinfo
   {journal} {Phys. Rev. E}\ }\textbf {\bibinfo {volume} {103}},\ \bibinfo
  {pages} {062133} (\bibinfo {year} {2021})}\BibitemShut {NoStop}%
\bibitem [{\citenamefont {Vidmar}\ \emph {et~al.}(2021)\citenamefont {Vidmar},
  \citenamefont {Krajewski}, \citenamefont {Bon\ifmmode~\check{c}\else
  \v{c}\fi{}a},\ and\ \citenamefont {Mierzejewski}}]{vidmar2021phenomenology}%
  \BibitemOpen
  \bibfield  {author} {\bibinfo {author} {\bibfnamefont {L.}~\bibnamefont
  {Vidmar}}, \bibinfo {author} {\bibfnamefont {B.}~\bibnamefont {Krajewski}},
  \bibinfo {author} {\bibfnamefont {J.}~\bibnamefont
  {Bon\ifmmode~\check{c}\else \v{c}\fi{}a}},\ and\ \bibinfo {author}
  {\bibfnamefont {M.}~\bibnamefont {Mierzejewski}},\ }\bibfield  {title}
  {\bibinfo {title} {Phenomenology of spectral functions in disordered spin
  chains at infinite temperature},\ }\href
  {https://doi.org/10.1103/PhysRevLett.127.230603} {\bibfield  {journal}
  {\bibinfo  {journal} {Phys. Rev. Lett.}\ }\textbf {\bibinfo {volume} {127}},\
  \bibinfo {pages} {230603} (\bibinfo {year} {2021})}\BibitemShut {NoStop}%
\bibitem [{Note4()}]{Note4}%
  \BibitemOpen
  \bibinfo {note} {In particular, other forms of comparison have the underlying
  assumption that the integrability-breaking parameters \(g\) and \(h\) can be
  treated similarly. Comparison of \(T_{E,\pm }/T_\lambda \) with \(T_{\lambda
  }/T^{*}_{\lambda }\) frees us of this assumption.}\BibitemShut {Stop}%
\bibitem [{\citenamefont {Karthik}\ \emph {et~al.}(2007)\citenamefont
  {Karthik}, \citenamefont {Sharma},\ and\ \citenamefont
  {Lakshminarayan}}]{karthik2007entanglement}%
  \BibitemOpen
  \bibfield  {author} {\bibinfo {author} {\bibfnamefont {J.}~\bibnamefont
  {Karthik}}, \bibinfo {author} {\bibfnamefont {A.}~\bibnamefont {Sharma}},\
  and\ \bibinfo {author} {\bibfnamefont {A.}~\bibnamefont {Lakshminarayan}},\
  }\bibfield  {title} {\bibinfo {title} {Entanglement, avoided crossings, and
  quantum chaos in an ising model with a tilted magnetic field},\ }\href
  {https://doi.org/10.1103/PhysRevA.75.022304} {\bibfield  {journal} {\bibinfo
  {journal} {Phys. Rev. A}\ }\textbf {\bibinfo {volume} {75}},\ \bibinfo
  {pages} {022304} (\bibinfo {year} {2007})}\BibitemShut {NoStop}%
\bibitem [{\citenamefont {Bulchandani}\ \emph {et~al.}(2022)\citenamefont
  {Bulchandani}, \citenamefont {Huse},\ and\ \citenamefont
  {Gopalakrishnan}}]{bulchandani2022onset}%
  \BibitemOpen
  \bibfield  {author} {\bibinfo {author} {\bibfnamefont {V.~B.}\ \bibnamefont
  {Bulchandani}}, \bibinfo {author} {\bibfnamefont {D.~A.}\ \bibnamefont
  {Huse}},\ and\ \bibinfo {author} {\bibfnamefont {S.}~\bibnamefont
  {Gopalakrishnan}},\ }\bibfield  {title} {\bibinfo {title} {Onset of many-body
  quantum chaos due to breaking integrability},\ }\href
  {https://doi.org/10.1103/PhysRevB.105.214308} {\bibfield  {journal} {\bibinfo
   {journal} {Phys. Rev. B}\ }\textbf {\bibinfo {volume} {105}},\ \bibinfo
  {pages} {214308} (\bibinfo {year} {2022})}\BibitemShut {NoStop}%
\bibitem [{\citenamefont {LeBlond}\ \emph {et~al.}(2021)\citenamefont
  {LeBlond}, \citenamefont {Sels}, \citenamefont {Polkovnikov},\ and\
  \citenamefont {Rigol}}]{leblond2021universality}%
  \BibitemOpen
  \bibfield  {author} {\bibinfo {author} {\bibfnamefont {T.}~\bibnamefont
  {LeBlond}}, \bibinfo {author} {\bibfnamefont {D.}~\bibnamefont {Sels}},
  \bibinfo {author} {\bibfnamefont {A.}~\bibnamefont {Polkovnikov}},\ and\
  \bibinfo {author} {\bibfnamefont {M.}~\bibnamefont {Rigol}},\ }\bibfield
  {title} {\bibinfo {title} {{Universality in the onset of quantum chaos in
  many-body systems}},\ }\href {https://doi.org/10.1103/PhysRevB.104.L201117}
  {\bibfield  {journal} {\bibinfo  {journal} {Phys. Rev. B}\ }\textbf {\bibinfo
  {volume} {104}},\ \bibinfo {pages} {L201117} (\bibinfo {year} {2021})},\
  \Eprint {https://arxiv.org/abs/2012.07849} {arXiv:2012.07849
  [cond-mat.stat-mech]} \BibitemShut {NoStop}%
\bibitem [{\citenamefont {LeBlond}\ \emph {et~al.}(2019)\citenamefont
  {LeBlond}, \citenamefont {Mallayya}, \citenamefont {Vidmar},\ and\
  \citenamefont {Rigol}}]{leblond2019entanglement}%
  \BibitemOpen
  \bibfield  {author} {\bibinfo {author} {\bibfnamefont {T.}~\bibnamefont
  {LeBlond}}, \bibinfo {author} {\bibfnamefont {K.}~\bibnamefont {Mallayya}},
  \bibinfo {author} {\bibfnamefont {L.}~\bibnamefont {Vidmar}},\ and\ \bibinfo
  {author} {\bibfnamefont {M.}~\bibnamefont {Rigol}},\ }\bibfield  {title}
  {\bibinfo {title} {Entanglement and matrix elements of observables in
  interacting integrable systems},\ }\href
  {https://doi.org/10.1103/PhysRevE.100.062134} {\bibfield  {journal} {\bibinfo
   {journal} {Phys. Rev. E}\ }\textbf {\bibinfo {volume} {100}},\ \bibinfo
  {pages} {062134} (\bibinfo {year} {2019})}\BibitemShut {NoStop}%
\bibitem [{\citenamefont {Brenes}\ \emph
  {et~al.}(2020{\natexlab{b}})\citenamefont {Brenes}, \citenamefont {Goold},\
  and\ \citenamefont {Rigol}}]{brenes2020low}%
  \BibitemOpen
  \bibfield  {author} {\bibinfo {author} {\bibfnamefont {M.}~\bibnamefont
  {Brenes}}, \bibinfo {author} {\bibfnamefont {J.}~\bibnamefont {Goold}},\ and\
  \bibinfo {author} {\bibfnamefont {M.}~\bibnamefont {Rigol}},\ }\bibfield
  {title} {\bibinfo {title} {Low-frequency behavior of off-diagonal matrix
  elements in the integrable xxz chain and in a locally perturbed
  quantum-chaotic xxz chain},\ }\href
  {https://doi.org/10.1103/PhysRevB.102.075127} {\bibfield  {journal} {\bibinfo
   {journal} {Phys. Rev. B}\ }\textbf {\bibinfo {volume} {102}},\ \bibinfo
  {pages} {075127} (\bibinfo {year} {2020}{\natexlab{b}})}\BibitemShut
  {NoStop}%
\bibitem [{Note5()}]{Note5}%
  \BibitemOpen
  \bibinfo {note} {A more accurate notion of the Lyapunov time would be to
  define it has the time at which the Krylov complexity becomes of the order of
  \(\protect \mathcal {K}\) (let's call it \(\tau _{*}\)). Here \(\protect
  \mathcal {K}\) is the saturation value of Krylov complexity. It is related to
  the dimension of the Krylov space (usually much less than that). The
  difference between the two choices is a factor of \(O(1)\) in the systems we
  study. So the ratio between \(\tau _{*}\) and \(T_\lambda =\alpha ^{-1}\)
  turns out to be \(O(1)\). Hence the study of \(\alpha ^{-1}\) is
  sufficient.}\BibitemShut {Stop}%
\bibitem [{\citenamefont {Fagotti}\ and\ \citenamefont
  {Essler}(2013)}]{fagotti2013reduced}%
  \BibitemOpen
  \bibfield  {author} {\bibinfo {author} {\bibfnamefont {M.}~\bibnamefont
  {Fagotti}}\ and\ \bibinfo {author} {\bibfnamefont {F.~H.~L.}\ \bibnamefont
  {Essler}},\ }\bibfield  {title} {\bibinfo {title} {Reduced density matrix
  after a quantum quench},\ }\href {https://doi.org/10.1103/PhysRevB.87.245107}
  {\bibfield  {journal} {\bibinfo  {journal} {Phys. Rev. B}\ }\textbf {\bibinfo
  {volume} {87}},\ \bibinfo {pages} {245107} (\bibinfo {year}
  {2013})}\BibitemShut {NoStop}%
\bibitem [{Note6()}]{Note6}%
  \BibitemOpen
  \bibinfo {note} {Corresponding to the off-diagonal fluctuation in
  ETH}\BibitemShut {NoStop}%
\end{thebibliography}%

\newpage

\clearpage

\newpage










%
\pagebreak

\begin{center}
    \textbf{\large Supplemental Material}
\end{center}

\setcounter{equation}{0}
\setcounter{figure}{0}
\setcounter{table}{0}
\setcounter{page}{1}
\setcounter{section}{0}

\makeatletter

\section{Krylov complexity}

The Krylov complexity of an operator \(\mop\) under the effect of the Hamiltonian \(H\), is computed with the following algorithm~\cite{parker2019A}, which generates a basis in the space of operators:
\begin{itemize}
    \item First element of the basis: \(\mop_{0} = \mop\).
    \item Evaluate the commutator of the operator with the Hamiltonian \(\mathcal{A}_1 = [H,\mop_0]\). 
    Note that this is orthogonal to \(\mop_0\).
    \item Normalize this operator \(\mop_{1} = \frac{1}{b_1}\mathcal{A}_{1}\) with \(b_1 = \sqrt{(A_1 \vert A_1)}\).
    This forms the second element \(\mop_1\) of the basis.
    \item The \(n^\mathrm{th}\) element of the Krylov basis is obtained by first evaluating
    \begin{align*}
        \mathcal{A}_{n} = [H,\mop_{n - 1}] - b_{n - 1}\mop_{n - 2}
    \end{align*}
    \item This \(\mathcal{A}_{n}\) is orthonogonal to all \(\mop_{k}\;\; \forall \;\; k < n\).
    \item Finally, normalize \(\mathcal{A}_{n}\) to obtain \(\mop_{n} = \frac{1}{b_n}\mathcal{A}_{n}\).
    This is the \(n^\mathrm{th}\) Krylov vector. 
    \item Terminate this process at \(\mathcal{K}\) where \(b_{\mathcal{K}} = 0\) and \(b_{\mathcal{K} - 1} > 0\).
\end{itemize}
This is a version of the Lanczos algorithm~\cite{viswanath1994recursion}.
The time-evolved operator \(\mot\) can now be written in the Krylov basis
\begin{align}
    \mot  = e^{i H t}\mop_{0}e^{-i H t} = \sum_{n = 0}^{\mathcal{K}}i^{n}\psi_{n}(t)\mop_{n}
\end{align}
The functions \(\psi_{n}(t)\) capture the time evolution of the operator \(\mop\).
Note that this algorithm rewrites the Baker-Campbell-Hausdorff expansion of \(\mot\) in a more compact form by essentially orthonormalizing each term with respect to all the others.
For the Hermitian initial operator (and Hamiltonian), \(i^{k}\mop_{k}\) is also Hermitian. 

The numbers \(b_{n}\) that have been collected from this algorithm uniquely fix all the functions \(\psi_{n}(t)\).
This is done by utilizing the fact that the autocorrelation function \(\psi_{0}(t) = (\mathcal{O}(t)\vert\mathcal{O}_{0})\) can be expanded in a Taylor series~\cite{viswanath1994recursion,parker2019A} of the form
\begin{align}
    \psi_{0}(t) = \sum_{k}\frac{\mu_{2 k}}{(2 k)!}t^{2 k}
\end{align}
where \( b_{1}^2 b_{2}^2 \dots b_{n}^2 = \text{det}(\mu_{i + j})_{0 \leq i, j \leq n} \). Once the function \(\psi_0\) is known, the remaining \(\psi_{k}\) can be figured by using the recursion relation
\begin{align}
    \partial_t \psi_{k}(t) = -b_{k + 1}\psi_{k + 1}(t) + b_{k}\psi_{k - 1}(t), \;\;\;\; \psi_{k}(0) = \delta_{k 0}
\end{align}
which follows from applying Heisenberg's equation on \(\mathcal{O}(t)\).

The sequence \(b_{n}\), called the Lanczos coefficients, can be used to distinguish between chaotic and integrable dynamics in certain cases.
The Operator Growth Hypothesis~\cite{parker2019A} states that chaotic dynamics is characterized by an (asymptotic) linear growth of the Lanczos coefficients, i.e. \(b_{n} \sim \alpha n\).
It can also be demonstrated the average position of the operator on this basis,
\begin{align}
    K(t) = \sum_{n}^{\mathcal{K}} n\vert \psi_{n}(t) \vert^2
\end{align}
called the \textit{Krylov complexity}, grows (asymptotically) as \(K(t) \sim e^{2 \alpha t}\) for chaotic systems. It is worth noting that this exponent also appears in the asymptotic decay rate of the spectral function $\Phi(\omega) = \int_{-\infty}^{\infty}\psi_0(t)e^{i \omega t}\mathrm{d}t$.
\begin{align}
    \Phi(\omega \rightarrow \infty) \sim e^{-\pi\vert\omega\vert/2\alpha}
\end{align}
The decay of the spectral function for large $\omega$ has been the subject of intense investigation in recent years and has been found to be a useful indicator of chaotic and integrable dynamics~\cite{vidmar2021phenomenology,leblond2021universality,leblond2019entanglement,brenes2020low,brenes2020eigenstate}.

A natural candidate for a Lyapunov exponent is the growth exponent \(\alpha\) of the Lanczos coefficients.
For systems that demonstrate chaotic dynamics, it has been argued~\cite{parker2019A} that the autocorrelation function \(\psi_0 (t)\) has poles on the imaginary axes and the lowest-lying one of those are given by \(t_0 = \pm \frac{\pi}{2\alpha}\).
Therefore the growth exponent \(\alpha\) can be extracted from the pole structure of the autocorrelation function. 

In the systems that we study in the main text, we present the behavior of \(\alpha\) (rather, the behavior of the \textit{Lyapunov time} \(T_\lambda=\alpha^{-1}\)) by choosing an appropriate initial operator \(\mop\)~\footnote{A more accurate notion of the Lyapunov time would be to define it has the time at which the Krylov complexity becomes of the order of \(\mathcal{K}\) (let's call it \(\tau_{*}\)). Here \(\mathcal{K}\) is the saturation value of Krylov complexity. It is related to the dimension of the Krylov space (usually much less than that). The difference between the two choices is a factor of \(O(1)\) in the systems we study. So the ratio between \(\tau_{*}\) and \(T_\lambda=\alpha^{-1}\) turns out to be \(O(1)\). Hence the study of \(\alpha^{-1}\) is sufficient.}.

{\color{black}
\section{Conserved Quantities: SRN}

When studying the SRN limit and computing the thermalization timescales, we choose the conserved quantity (in the integrable limit $h \rightarrow 0$) to be the Pauli matrix $\sigma^{z}_{i}$, at some lattice site $i$. This quantity is local, and so the integrability breaking induces extra terms that are also local (but not necessarily $1-$ local).
It is interesting to consider what would happen if a \emph{non-local} conserved quantity is instead considered in the SRN case.
The results do not change much for the following reasons: Let us consider the non-local initial operator
\begin{align}
    \mop{(N)} = \sum_{i = 1}^{N}\sigma^{z}_{i}
\end{align}
where the superscript \((N)\) is used to indicate that the operator is non-local.
Correspondingly, the local operator is denoted as \(\mop^{(1)} = \sigma^{z}_{1}\).
The time evolution of the operator \(\) can be broken into that of individual \(\sigma^{z}_{i}\).
The evolution for each of these should be equivalent since the state (with respect to which the expectation value is calculated) \(\ket{\psi}\) is a random state and hence has roughly equal weight at each site \(i\).
This allows us to approximate the fluctuation equation as
\begin{align}
    \langle\mop^{(N)}(t)\rangle - \overline{\mop^{(N)}} \approx N\left(\langle\mop^{(1)}(t)\rangle - \overline{\mop^{(1)}}\right)\,,
\end{align}
which has a similar distribution of zeros, and hence the similar moments, as that of \(\mop^{(1)}\). 

The Krylov complexity (or Lanczos growth) of such non-local operators should also be the same as that of local operators.
This is because we are working with translation symmetric systems, and therefore each \(\sigma_i\) can be replaced by \(\sigma_1\) in the BCH expansion.
The resulting overall factor of \(N\) is taken care of via normalization.

To support this argument, we present the numerical results for the two cases in Fig.~\ref{fig:compsum}.
The non-local timescales are represented by a superscript \((N)\).
    
\begin{figure}
    \centering
    \includegraphics[width=0.95\columnwidth]{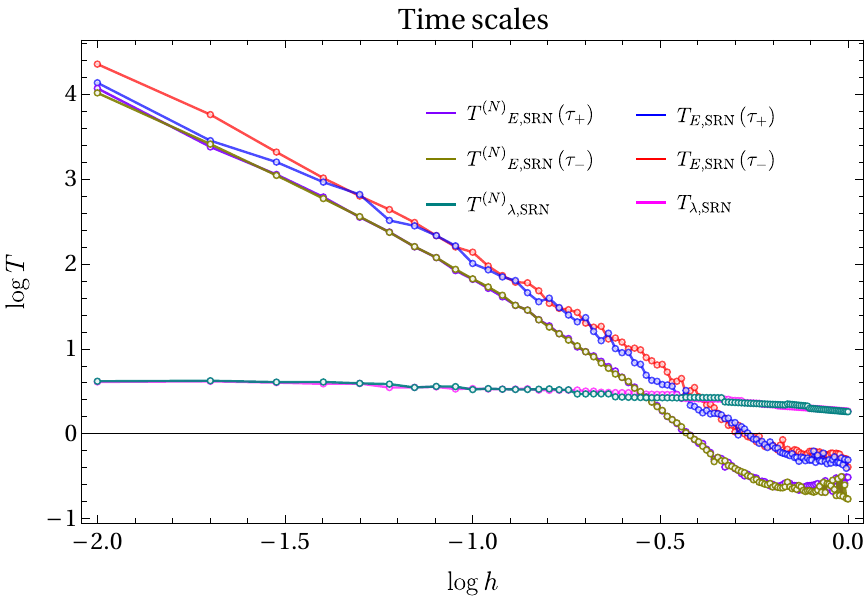}
    \caption{
        Comparison of ergodization times and Lyapunov times for SRN with local and non-local operators.
        The superscript \((N)\) stands for the non-local operator.
        The initial states are different but chosen from the same random distribution, for the local and non-local cases.
    }
    \label{fig:compsum}
\end{figure}

It is interesting to note that the ergodization timescales for \(\tau_{\pm}\) are much more similar for the case of \(\mop^{(N)}\) than for \(\mop^{(1)}\).
This is due to the sum of local operators having a ``smoothing'' effect on the random state \(\ket{\psi}\).
This causes the timescales obtained from \(\tau_{+}\) and \(\tau_{-}\) to almost exactly overlap in the weakly integrable limit.

\section{Conserved Quantities: LRN}
Here we discuss the conserved quantities of the integrable limit $g\rightarrow0$ of the long-range network class. This integrable Hamiltonian is known as the transverse field Ising model and has a complete set of conserved quantities. These are given by
\begin{align}
    I^{(k)} &= i J \sum_{j = 1}^{N}\left(S^{z y}_{j:j+k} - S^{y z}_{j:j+k}\right)\;\;\;\; k = 1,\dots,N-1
\end{align}
where we have the following shorthand
\begin{align}
    S^{\alpha \beta}_{j:j+l} = \sigma^{\alpha}_{j}\left(\prod_{n = 1}^{l - 1}\sigma^{x}_{j+n}\right)\sigma^{\beta}_{j + l}
\end{align}
It is straightforward to see that $[H_{\text{TFIM}}, I^{(k)}] = 0$. These quantities can be interpreted as linear combinations of mode occupation numbers in the Jordan-Wigner fermion theory \cite{fagotti2013reduced}. For $k = N$, the conserved quantity corresponds to $\prod_{i = 1}^{N}\sigma^{x}_{i}$, which is a symmetry operation corresponding to the replacement $\sigma^{z,y} \rightarrow -\sigma^{z,y}$. 
\begin{figure}
    \centering
    \includegraphics[width=0.95\columnwidth]{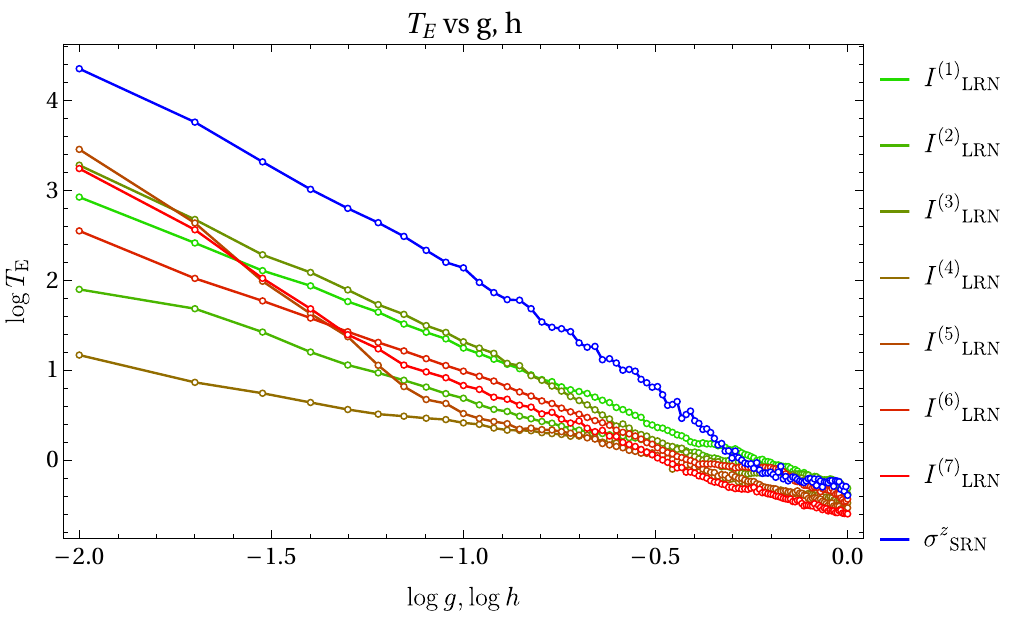}
    \caption{
        Scaling of ergodization times for LRN (for all conserved quantities $I^{(k)}$) and comparison with the same for SRN (single operator). 
    }
    \label{fig:allI}
\end{figure}

We compare the ergodization times $T_{E,\pm}$ for each $I_{k}$, with the respective Lyapunov time $T_{\lambda}$ obtained via the Krylov method. The results are presented in Fig.~\ref{fig:allI}. It is evident that the scaling is different for different $I^{(k)}$. Note that despite different scaling behaviour of the ergodization time $T_{E}$ for different conserved quantities $I^{(k)}$, the ergodization time for the $\sigma^{z}$ operator in the SRN case bounds the $T_{E}$ for all $I^{(k)}$ from above. Assuming a power law scaling behavior for the timescales
\begin{align}
    \log T_{E, \lambda} = \begin{cases}
    &\alpha_{E,\lambda} \log g + \delta_{E,\lambda} \;\;,\;\; \text{LRN}\\
    &\alpha_{E,\lambda} \log h + \delta_{E,\lambda} \;\;,\;\; \text{SRN}
    \end{cases}
\end{align} we obtain the exponents listed in Table~\ref{tab:pwlaw}.
\begin{table}[]
    \centering
    \begin{tabular}{|c|c|c|c|c|}
        \hline
        $k$ & $\alpha^{+}_{E}$ & $\delta^{+}_{E}$ & $\alpha_{\lambda}$ & $\delta_{\lambda}$\\
        \hline
        $1$ & $-1.65828$ & $-0.395546$ & $-0.046659$ & $0.262234$\\
        \hline
        $2$ & $-1.28645$ & $-0.592328$ & $-0.0204276$ & $0.248453$\\
        \hline
        $3$ & $-1.9492$ & $-0.638931$ & $-0.151105$ & $0.00884637$\\
        \hline
        $4$ & $-0.734752$ & $-0.356261$ & $-0.169417$ & $-0.0685405$\\
        \hline
        $5$ & $-3.03973$ & $-2.60729$ & $-0.154005$ & $-0.0406314$\\
        \hline
        $6$ & $-1.54398$ & $-0.570509$ & $-0.155603$ & $-0.0419308$\\
        \hline
        $7$ & $-2.49685$ & $-1.76273$ & $-0.196703$ & $-0.11035$\\
        \hline
    \end{tabular} 
    \caption{Power-law coefficients for all $I^{(k)}$ for the LRN case. The superscript $+$ stands for results extracted from positive passage times. The results for negative passage times are comparable.}
    \label{tab:pwlaw}
\end{table}
For the corresponding SRN case, we observe that the coefficients are
\begin{align}
    &\alpha^{+}_{E, SRN} = -2.26301 \;\;,\;\; \delta^{+}_{E, SRN} = -0.136552 \notag\\
    &\alpha_{\lambda, SRN} = -0.0817788\;\;,\;\; \delta_{\lambda, SRN} = 0.217633 \notag
\end{align}
It is also instructive to compare the scaling of the ratio of timescales $T_{E}/T_{\lambda}$ as a function of $T_{\lambda}$. This ratio is expected to diverge upon approaching the integrable limit (i.e. as $T_{\lambda}$ increases). Since different conserved quantities will in general have different ranges of values of $T_{\lambda}$ (although within the same order), it is better to instead study the ratio as a function of $T_{\lambda}/\text{max}(T_{\lambda})$. We present this result in Fig.~\ref{fig:all2}.
\begin{figure}
    \centering
    \includegraphics[width=0.95\columnwidth]{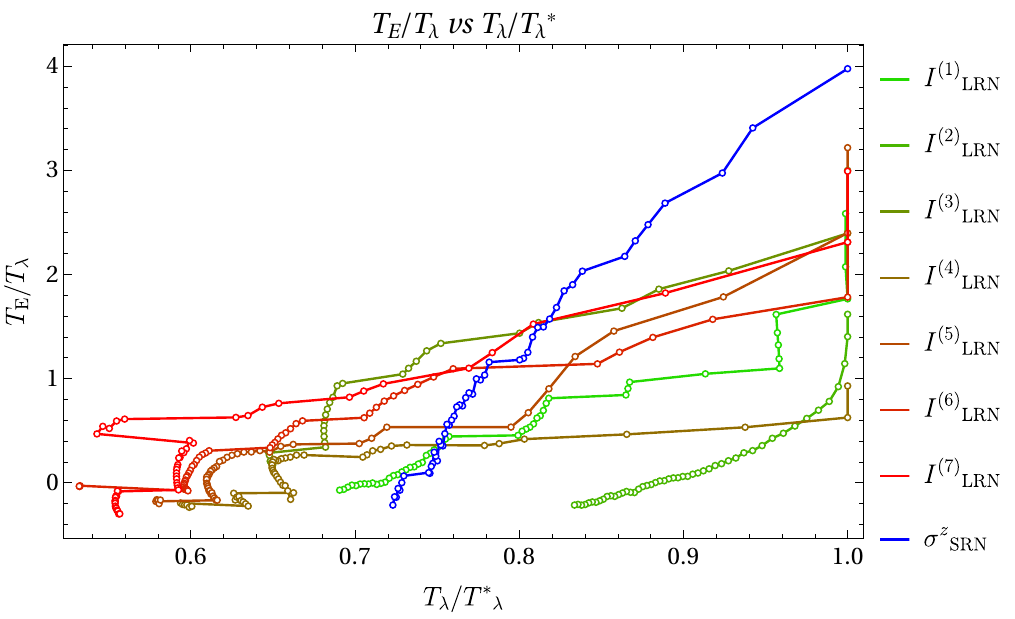}
    \caption{
        Behavior of the ratio $T_{E}/T_{\lambda}$ as a function of $T_{\lambda}/T^{*}_{\lambda}$, where $T^{*}_{\lambda} = \text{max}(T_{\lambda})$. The scaling for all $I^{(k)}$ (LRN) is compared to that of $\sigma^{z}$ (SRN). These results are presented for the integrability-breaking parameter close to $0$.
    }
    \label{fig:all2}
\end{figure}

Finally, we consider the scaling of $T_{\lambda}$ and $T_{E}/T_{\lambda}$ with the integrability-breaking parameter $g, h$ for completeness. This is presented in Fig.~\ref{fig:all3} and Fig.~\ref{fig:all4} respectively.
\begin{figure}
    \centering
    \includegraphics[width=0.95\columnwidth]{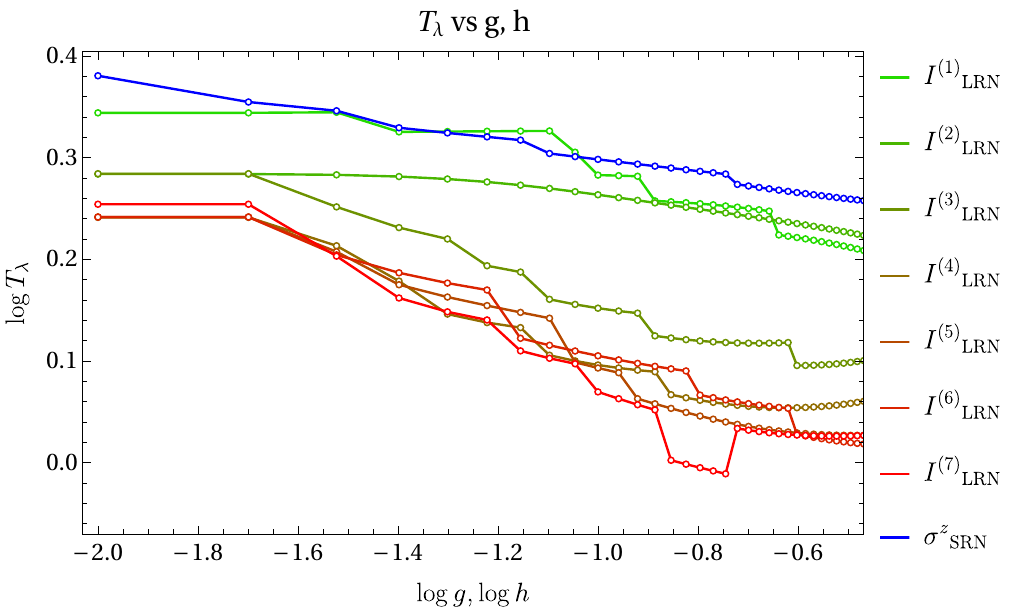}
    \caption{
        Behavior of the Lyapunov times $T_{\lambda}$ as a function of the integrability breaking parameter $g, h$. The region close to the integrable limit is explored. It is observed that for all conserved quantities $I^{(k)}$ (LRN), as well as for the SRN observable $\sigma^{z}$, the Lyapunov times are comparable.
    }
    \label{fig:all3}
\end{figure}
\begin{figure}
    \centering
    \includegraphics[width=0.95\columnwidth]{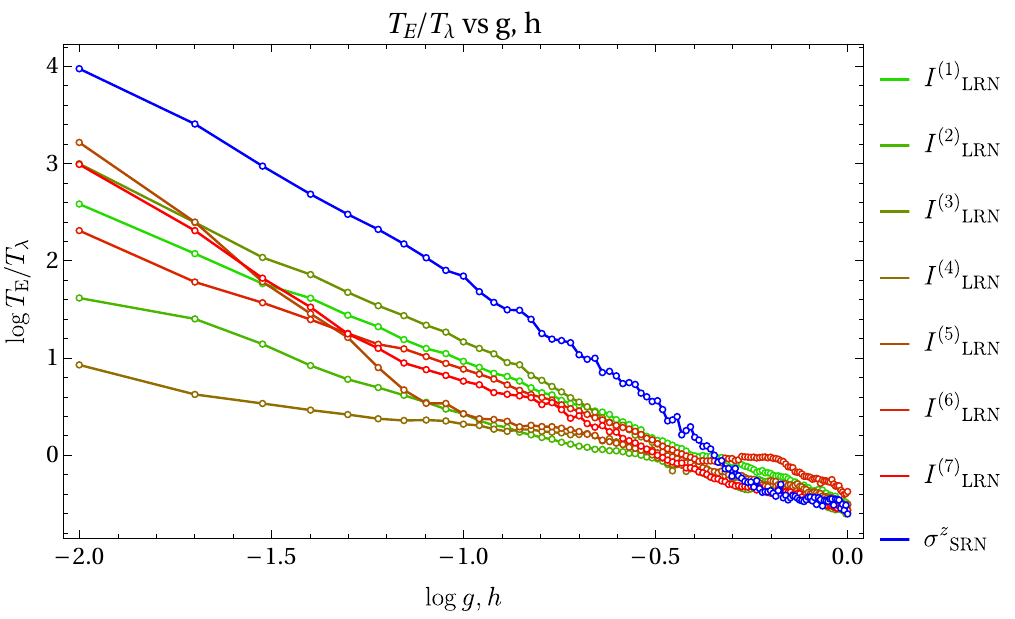}
    \caption{
        Behavior of the ratio $T_{E}/T_{\lambda}$ as a function of the integrability breaking parameter $g, h$. The results support the conclusion of Fig.~\ref{fig:allI}.
    }
    \label{fig:all4}
\end{figure}
The results of Fig.~\ref{fig:all3} and \ref{fig:allI} explain the observation in Fig.~\ref{fig:all4}, since the Lyapnuov times $T_{\lambda}$ for all $I^{(k)}$ (LRN) and $\sigma^{z}$ (SRN) remain comparable  throughout the range of $g, h$ explored. However, the ergodization times scale in a different manner (significantly different, on the log-scale, as seen in Fig.~ \ref{fig:allI}). Thus the ratio $T_{E}/T_{\lambda}$ is also highly sensitive to initial operator choice and the universality classification.
}

\section{Passage times}

\begin{figure}
    \centering
    \includegraphics[width=0.95\columnwidth]{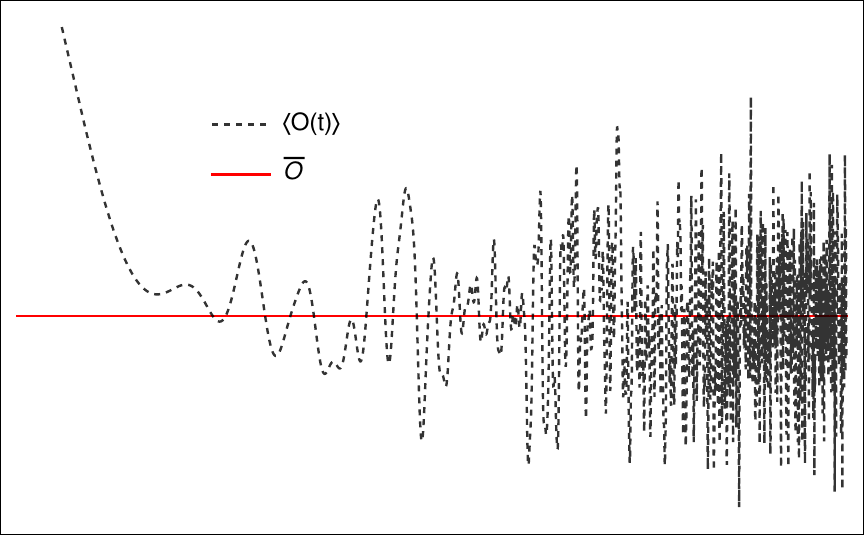}
    \caption{
        Schematic representation of time-evolution of \(\langle \mot \rangle\) around the mean value \(\omo\).
        We have used a log scale on the \(t\)-axis. 
    }
    \label{fig:pt}
\end{figure}

The expectation value of an operator \(mop\) oscillates around its mean value at long times.
Consider a Hamiltonian \(H\) and some operator \(\mop\) whose time evolution is studied under this Hamiltonian.
This is schematically shown in the Fig.~\ref{fig:pt}.
The time evolved operator is given by
\begin{align}
    \mot = e^{-i H t}\mot e^{i H t}
\end{align}
The expectation value of this operator in a generic state \(\ket{\psi}\) is given by
\begin{align}
    \langle \mot \rangle = \mel{\psi}{e^{-i H t}\mop e^{i H t}}{\psi}
\end{align}
The generic state is written as follows in terms of the eigenstates \(\ket{n}\) of the Hamiltonian \(H\)
\begin{align}
    \ket{\psi} = \sum_{n}c_{n}\ket{n}
\end{align}
The expectation value $\langle \mot \rangle$ can now be rewritten as
\begin{align}
    \langle \mot \rangle  = \sum_{m, n}c_{n}c^{*}_{m}e^{i (E_{n} - E_{m})t}\mel{m}{\mathcal{O}}{n}
\end{align}
The time-averaged value of this expectation is given as
\begin{align}
    \bar{\mop} &= \lim_{T \rightarrow \infty}\frac{1}{T}\int_{0}^{T}\langle \mot \rangle \mathrm{d}t  \\
    &= \lim_{T \rightarrow \infty}\sum_{m, n}c_{n}c^{*}_{m}\mel{m}{\mop}{n}\left(\frac{1}{T}\int_{0}^{T}e^{i(E_{n} - E_{m})t}\mathrm{d}t\right)
\end{align}
The integral over \(t\) gives a \(\delta(E_{n} - E_{m})\).
Therefore the final result is
\begin{align}
    \bar{\mop} = \sum_{n}\vert c_{n} \vert^2 \mel{n}{\mop}{n} + \sum_{n', m'} c_{n'} c^{*}_{m'} \mel{m'}{\mop}{n'}
\end{align}
where the second sum is over all \(n', m'\) for which \(E(n') = E(m')\).
So if the mean value is subtracted from \(\langle \mot \rangle\), we obtain
\begin{align}
    &f_{\mop}(t) = \langle \mot \rangle  - \bar{\mop}\notag\\
    &= \sum_{m, n - \{m', n'\}} c_{n} c^{*}_{m} e^{i (E_{n} - E_{m})t} \mel{m}{\mop}{n} - \sum_{n}\vert c_{n} \vert^2 \mel{n}{\mop}{n}
    \label{foeqn}
\end{align}
where the terms corresponding to degeneracies were dropped.
This captures the behavior of the off-diagonal elements of the time-evolved operator.

We evaluate the distribution of the zeros of the function \( f_{\mop}(t)\) and determine how they are spaced.
The moments of the distribution of this spacing can be interpreted as another natural time scale.
Note that for random uniform initial state (i.e. \(c_{n}\) are uniform random numbers), this distribution is determined by the level spacing distribution of the Hamiltonian and the off-diagonal elements of the initial operator~\footnote{Corresponding to the off-diagonal fluctuation in ETH}.

The passage or excursion times are then defined as the interval \(\tau_{i}\) between the zeros \(t_{i}\) and \(t_{i + 1}\) of the function \(f_{\mop}(t)\).
There are two passage times which are extracted from this information.
The first is the positive passage time \(\tau_{i,+}\) which corresponds to \(f_{\mop}(t)\) being positive in the interval \(t_{i}\) to \(t_{i + 1}\).
The negative passage time \(\tau_{j,-}\) corresponds to \(f_{\mop}(t)\) being negative in the interval \(t_{j}\) to \(t_{j + 1}\).
In this manuscript, we study the statistical distribution of \(\tau_{i, \pm}\) through their mean and variance (and combinations thereof).

\section{Details of the numerics}

In this section, we discuss the details of the numerical computations and present some of the results that are mentioned in the main text. 

While determining the passage times, our approach involved first diagonalizing the Hamiltonian \(H\)~\eqref{eq:TFIM} to find its eigenvalues and eigenvectors.
The next step is determining the coefficients \(c_{n}\) corresponding to the initial state \(\psi\), drawn from a uniform distribution, and the components of the initial operator \(\mel{m}{\mop}{n}\) and then plugging it in the expression~\eqref{foeqn}.
Then this function was evaluated numerically by varying \(t\) in steps of \(t_0 = \min_{m \neq n}\frac{1}{4(E_m - E_n)}\) and the values \(t = t_{i}\) for which \(f(t_{i}) = 0\) were collected.
Finally the difference \(\tau_{i} = t_{i + 1} - t_{i}\) was computed to determine the excursion times.
We collected \(\sim 10^4\) passages.

\begin{figure}
    \centering
    \begin{subfigure}{0.48\textwidth}
        \includegraphics[width=\textwidth]{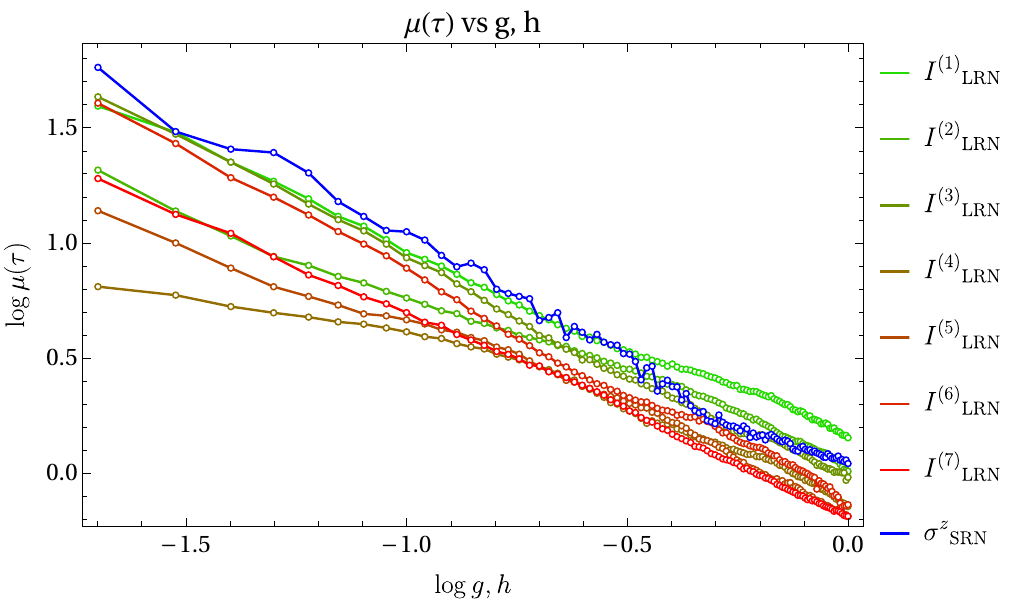}
        \caption{
            Mean of ergodization times \(\mu(\tau)\).
        }
        \label{fig:111}
    \end{subfigure}
    \begin{subfigure}{0.48\textwidth}
        \includegraphics[width=\textwidth]{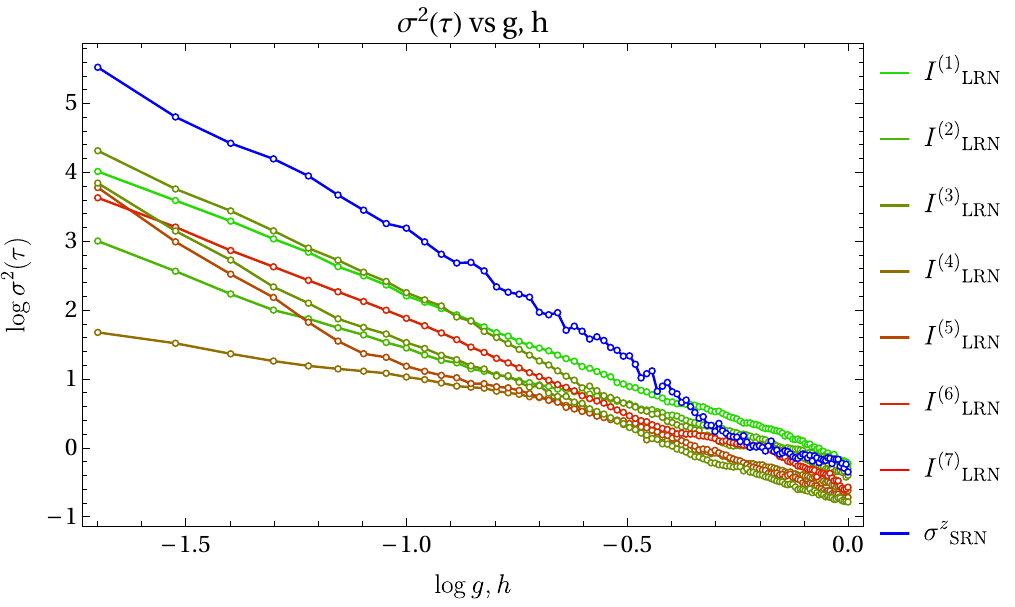}
        \caption{
            Variance of ergodization times \(\sigma^2(\tau)\).
        }
        \label{fig:122}
    \end{subfigure}
    \caption{
        Mean and Variance of ergodization times for all operators \(I^{(k)}\) and the SRN operator \(\sigma^{z}\), as a function of $g$ and $h$ respectively.
    }
    \label{fig:momentsall}
\end{figure}

\begin{figure}
    \centering
    \begin{subfigure}{0.41\textwidth}
        \includegraphics[width=\textwidth]{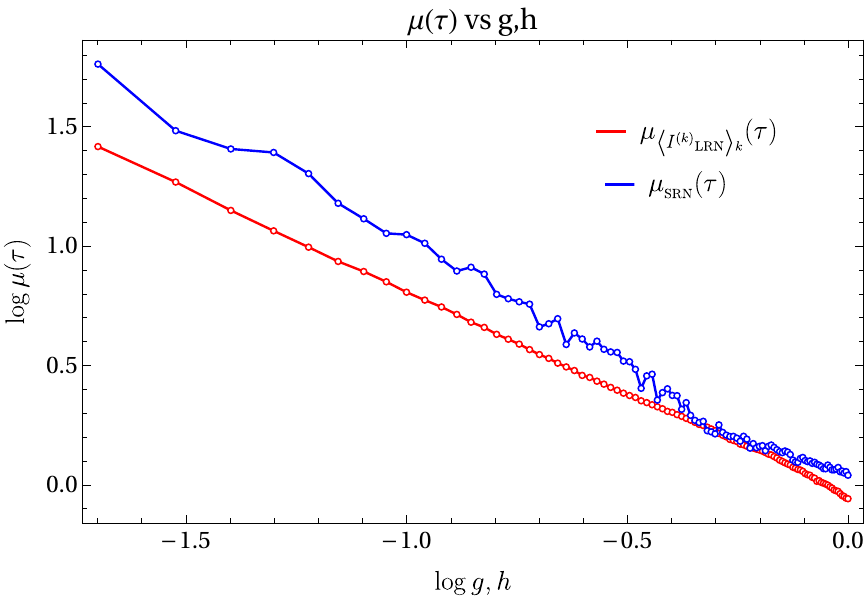}
        \caption{
            Mean of ergodization times \(\mu(\tau)\).
        }
        \label{fig:111}
    \end{subfigure}
    \begin{subfigure}{0.41\textwidth}
        \includegraphics[width=\textwidth]{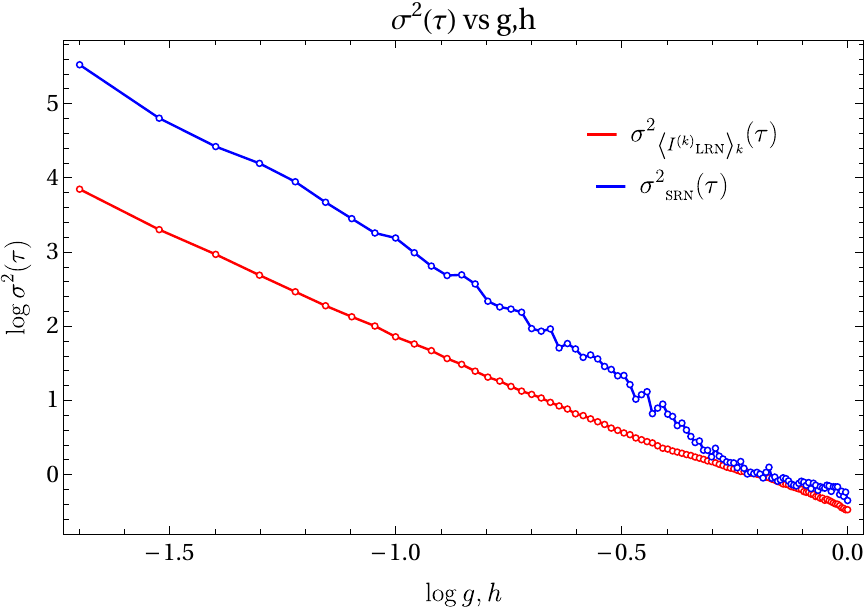}
        \caption{
            Variance of ergodization times \(\sigma^2(\tau)\).
        }
        \label{fig:122}
    \end{subfigure}
    \caption{
        Mean and Variance of ergodization times for the averaged operator \(N^{-1} \sum_{k} I^{(k)}\) and the SRN operator \(\sigma^{z}\), as a function of $g$ and $h$ respectively.
    }
    \label{fig:momentsavg}
\end{figure}

The results for the short- and long-range networks are discussed in the main text.
For the short- and long-range network, the appropriate choice of ergodization time is the ratio of variance and mean of the excursion times as supported by the data shown in Fig.~\ref{fig:momentsall} and Fig.~\ref{fig:momentsavg}.
The exponentially larger scale of \(\sigma^2\) as compared to \(\mu\) suggests that the fluctuations dominate the dynamics. Therefore, the appropriate choice of a timescale would be a ratio of the fluctuation to the mean, given by \(\frac{\sigma^2}{\mu}\). This is found to be the case in both LRN and SRN.

\end{document}